\documentclass[aps,pra,twocolumn,showkeys,nofootinbib,superscriptaddress]{revtex4}
\pdfoutput=1

\usepackage[utf8]{inputenc}
\usepackage{graphicx}
\usepackage{graphics}
\usepackage{epsfig}
\usepackage{lscape}
\usepackage{color}
\usepackage{multirow}
\usepackage{hyperref}
\usepackage{breqn}
\usepackage{physics}
\usepackage{subfiles}

\newcommand{\HIJ}{
Helmholtz-Institut Jena, 
Fr\"o{}belstieg 3, D-07743 Jena, Germany \\
}
\newcommand{\GSI}{
GSI Helmholtzzentrum f\"ur Schwerionenforschung,
Planckstrasse 1, 64291 Darmstadt, Germany\\
}
\newcommand{\Chemie}{
Department Chemie,
Johannes Gutenberg-Universit\"at, Fritz-Strassmann Weg 2, 55128 Mainz, Germany \\
}
\newcommand{\HIM}{
Helmholtz-Institut Mainz,
Staudingerweg 18, 55128 Mainz, Germany \\
}
\newcommand{\RUG}{
Van Swinderen Institute for Particle Physics and Gravity,
University of Groningen, Nijenborgh 4, 9747 Groningen, The Netherlands\\
}
\newcommand{\US}{
Science Department, Chatham University, Pittsburgh, Pennsylvania 15232, USA\\
}
\newcommand{\JGU}{
Institut f\"ur Physik, Johannes-Gutenberg Universit\"at Mainz, Staudingerweg 7, 55128 Mainz, Germany\\
}
\newcommand{\FSU}{
Theoretisch-Physikalisches Institut, Friedrich-Schiller-Universit\"at Jena, D-07743 Jena, Germany\\
}
\newcommand{\KIT}{
Karlsruhe Institute of Technology (KIT), Institute for Nuclear Waste Disposal (INE), D-76021 Karlsruhe, Germany
}

\begin{document}
\title{Transport property predictions for laser resonance chromatography on Rf$^+$ (Z = 104)}
\author{Giorgio Visentin\footnote{Corresponding author: g.visentin@hi-jena.gsi.de}}
\affiliation{\HIJ} \affiliation{\GSI}
\author{Harry Ramanantoanina\footnote{Corresponding author: harry.ramanantoanina@kit.edu}}
\affiliation{\Chemie} \affiliation{\HIM} \affiliation{\KIT}
\author{Anastasia Borschevsky}
\affiliation{\RUG} 
\author{Larry Viehland}
\affiliation{\US}  
\author{Biswajit Jana}
\affiliation{\Chemie}
\affiliation{\HIM}
\author{Aayush Arya}
\affiliation{\HIM}
\affiliation{\JGU}
\author{Stephan Fritzsche}
\affiliation{\HIJ}
\affiliation{\GSI}
\affiliation{\FSU}
\author{Mustapha Laatiaoui}
\affiliation{\Chemie} \affiliation{\HIM}
\date{\today}

\begin{abstract}
We propose a theoretically designed laser resonance chromatography (LRC) experiment on Rf$^+$ (Z = 104) drifting in He buffer gas. To this end, we first developed a four-level rate equation model that simulates the optical pumping of Rf$^+$ from its ground state, $^2$D$_{3/2}$ (7s$^2$6d$^1$), to the metastable $^4$F$_{3/2}$ (7s$^1$6d$^2$) state via laser resonant excitation of the intermediate $^4$F$_{3/2}$ (7s$^1$6d$^1$7p$^1$) state prior to electronic state chromatography. This model predicts a 93\% pumping efficiency that suffices to enable efficient laser resonance chromatography of this ion. We then performed accurate relativistic Multi-Reference Configuration-Interaction (MRCI) calculations to model the interaction of Rf$^+$ with He in the ground $^2$D$_{3/2}$ (7s$^2$6d$^1$), low-lying $^2$D$_{5/2}$ (7s$^2$6d$^1$), and metastable $^4$F$_{3/2}$ (7s$^1$6d$^2$) states. These ion-atom interaction potentials were used to calculate the state-specific ion mobilities. For gas temperatures above 100 K and small applied electric fields, the reduced ion mobilities of the ground and metastable states differ significantly. In particular, at room temperature the difference between the reduced ion mobilities of these states is larger than 11\%, and \textcolor{black}{as such sufficiently large to ensure LRC of this ion}.  
\end{abstract}
\maketitle

\section{Introduction}
The nuclides with atomic number Z $>$ 103 are known as the $Superheavy$ $elements$ \cite{Nazarewicz:2018, Dullmann:2017, Ramanantoanina:2021}. Such species are radioactive and produced synthetically in nuclear laboratories \cite{Nazarewicz:2018,Flerov:1981}. The scientific community has become interested in superheavy elements since their first synthesis in the 1960s \cite{Flerov:1983, Laatiaoui:2019}. Multiple questions drive this interest; for instance, the search for the nuclear stability island in the atomic number Z region between 114 and 126 and neutron number N = 184 \cite{Laatiaoui:2019, Oganessian:2015, Dullmann:2017}, alongside their nuclear \cite{Asai:2015}  and electronic \cite{Visentin:2020, Ramanantoanina:2021} structures, which are in turn affected by strong relativistic and quantum electrodynamics effects \cite{Rickert:2020}.  \\
The synthesis and characterization of these elements are extremely challenging; in fact, they can be produced only in a one-atom-per-second regime \cite{Ramanantoanina:2021, Visentin:2020, Laatiaoui:2019,Laatiaoui:2020a, Dullmann:2017, Sewtz:2003} and their production cross-section rapidly decreases with Z \cite{Nazarewicz:2018, Visentin:2020, Flerov:1983}. Furthermore, they are short-lived \cite{Ramanantoanina:2021, Dullmann:2017}. Standard spectroscopic experiments based on fluorescence detection suffer from low sensitivity and therefore cannot be exploited to characterize short-living species with such low production rates \cite{Visentin:2020, Campbell:2016}. Thus, novel techniques based on gaseous transport properties are explored in order to determine the elusive electronic structure and properties of superheavy elements \cite{Visentin:2020, Rickert:2020}. An example of these novel techniques is Laser Resonance Chromatography (LRC) \cite{Laatiaoui:2020a}, that is specifically developed for the characterization of superheavy elements. LRC exploits optical laser-pumping of ions drifting in dilute buffer gases (e.g.\textcolor{black}{helium}), in order to detect optical resonances. Ions in distinct electronic states experience different interactions with the buffer gas and, therefore, under the influence of an external and homogeneous electric field, move through the drift tube with different velocities toward the particle detector \cite{Ramanantoanina:2021, Ramanantoanina:2023, Laatiaoui:2020a, RomeroRomero:2022}. The interrelationship between the electronic state of the ion and its gas-phase transport properties enables the state-specific ion separation and resonance detection, according to a phenomenon called the $electronic$-$state$ $chromatographic$ (ESC) effect \cite{Visentin:2020, Buchachenko:2022}. \\
Thus far, a close interplay of experiments and electronic structure calculations has allowed the successful validation of this technique for several singly-charged lanthanide and actinide ions in the ground state \cite{Visentin:2020, Ramanantoanina:2023, Ramanantoanina2022, Kahl:2019, Laatiaoui:2012, Manard:2017}. For a few of them (see for instance \cite{Ramanantoanina:2023}) this interplay even helped determine the optimal experimental conditions to ensure that the transport properties of the ground state substantially differ from those of the metastable state. In contrast, no study of this kind exists for superheavy ions.   \\
The validation of LRC for superheavy elements should indeed start with Rf$^+$. \textcolor{black}{Rf$^+$ has Z = 104 and is, therefore, the first of the superheavy elements in the periodic table}. The electronic structure of this ion is non-trivial, as it is affected by remarkable electron correlation and strong relativistic effects \cite{Ramanantoanina:2021, Indelicato:2011, eliav1995, Fritzsche:2022}. In spite of this complexity, studies dating back to the 1990s have been devoted to the investigation of its energy levels \cite{Ramanantoanina:2021, Allehabi:2021, Martin:1996}, atomic radii \cite{Jonhson:1990}, ionization potentials \cite{Jonhson:1990}, oxidation states and chemical properties \cite{Pershina:1994, Liu:1999, Anton:2003, Pershina:2014}. These studies showed that Rf$^+$ features a few well-spaced long-living excited states that can be exploited for the optical-pumping step of the LRC experiment; in this regard, Ramanoantoanina $et$ $al.$ \cite{Ramanantoanina:2021} proposed a four-level optical pumping scheme. This consists of the ground state $\ket{1}$: $^2$D$_{3/2}$ (7s$^2$6d$^1$), the intermediate state $\ket{2}$: $^4$F$_{3/2}$ (7s$^1$6d$^1$7p$^1$), at about 28000 cm$^{-1}$ \cite{Ramanantoanina:2021}, that should be probed by laser radiation, the metastable state $\ket{3}$: $^4$F$_{3/2}$ (7s$^1$6d$^2$) at about 15000 cm$^{-1}$, which serves to collect the population from the $\ket{2}$ by radiative relaxation processes, and the low-lying state $\ket{4}$: $^2$D$_{5/2}$ (7s$^2$6d$^1$) between 5682 and 7444 cm$^{-1}$ \cite{Allehabi:2021, Ramanantoanina:2021}. \\
In this work, we start from the optical pumping scheme outlined above and perform accurate relativistic $ab$ $initio$ computations to devise and validate an LRC experiment on Rf$^+$. We first develop a rate equation model based on this optical pumping scheme and evaluate its efficiency. Then, we compute the ion-atom interaction potentials and ion mobilities for Rf$^+$interacting with He in the ground state, $^2$D$_{3/2}$ (7s$^2$ 6d$^1$), and the two corresponding longest-living excited states taken into account by the rate equation model, i.e., the low-lying $^2$D$_{5/2}$ (7s$^2$ 6d$^1$) state and the metastable state $^4$F$_{3/2}$ (7s$^1$ 6d$^2$). The computed ion mobilities are finally used to determine the experimental settings that ensure the optimal state-specific discrimination of the Rf$^+$ gas-phase transport properties. Indeed, the goal of this theoretical investigation is to provide a strategy for prospective LRC experiments aimed at the characterization of superheavy elements. Our study is organized as follows; we   detail the computational approach in Section II, while Section III is devoted to discussing $i$) the rate equation model based on the optical pumping scheme, $ii$) the ion-atom interaction potentials, and $iii$) the ion mobilities and their state-specific differences in terms of several experimental settings. Finally, we report the conclusions of our investigation in Section IV.     \par

\section{Methodology and Computational Details}
\subsection{Ion-atom interaction potentials}
The \textit{ab initio} Multi-Reference Configuration-Interaction (MRCI) calculations of the Rf$^+$-He interaction potentials ($V(d)$) were performed using the DIRAC19 code \cite{DIRAC19}. The calculations were carried out in the framework  of the four component Dirac-Coulomb Hamiltonian, and the nuclei were treated within a finite-nucleus model \textit{via} the Gaussian charge distribution \cite{VisDya97}. The uncontracted Gaussian-type Dyall basis sets \cite{dyall2004,dyall2011} of single-augmented triple-zeta (s-aug-v3z) quality were used for all the elements. \textcolor{black}{Note that the choice of basis set quality influences the theoretical results. Previous work on MRCI calculation of Lr+ electronic structure with complete basis set limit analysis revealed that only small differences are obtained with triple-zeta and quadruple-zeta basis set.\cite{Ramanantoanina2022}} The metal ion and the neutral helium atom were placed along the \textit{z}-axis in a system of Cartesian coordinates, separated by an inter-atomic distance $d$ that was varied from $2.0\,$\AA~ to $40.0\,$\AA~ for the calculation of the interaction potentials. We used the Boys-Bernardi counterpoise correction to tackle basis set superposition error \cite{Boys1970}: $V(d)=E_{Rf^+-He}(d)-E_{Rf^+}(d)-E_{He}(d)$. $E_{Rf^+-He}(d)$ is the MRCI energy of the M$^+$-He system at an inter-atomic distance $d$. $E_{Rf^+}(d)$ and $E_{He}(d)$ are the energies of the systems Rf$^+$-Gh and Gh-He, respectively, where He and Rf atoms are replaced by a ghost atom (Gh) without charge but carrying the the full basis sets of the He and Rf elements, respectively.\par
The electronic structure was obtained in two steps. In the first step, Dirac-Hartree-Fock calculations were performed using the average of configuration (AOC) type calculation. The AOC allowed us to represent the open-shell electronic structure system with 3 valence electrons that were evenly distributed over 12 valence spinors (6 Kramers pairs) of 7\textit{s} and 6\textit{d} atomic characters. The resulting wavefunction was used as reference for the CI calculations.  In the second step, the energy levels and the spectroscopic properties were calculated using the MRCI approach, within the Kramers-restricted configuration interaction module in DIRAC19\cite{DIRAC19,saue2020,thyssen2008,knecht2010}. In this implementation, the Kramers-restricted configuration interaction calculations use the concept of generalized active space (GAS) \cite{fleig2003}, which enables MRCI calculations with single and double electron excitations for different GAS set-ups \cite{saue2020}. The MRCI model \textit{a priori} takes into consideration the dynamical correlation of the active electrons \cite{fleig2012}. \par  
\begin{table}[htp]
\centering
\caption{Specification of the generalized active space (GAS) scheme used in the calculations of the Rf$^+$-He systems. See text for details.}
\label{table1}
\begin{tabular}{ l| c| c| c| c}
\hline\hline
GAS       & \multicolumn{2}{|c|}{Accumulated}  & Number of & Characters \\
Space     & \multicolumn{2}{|c|}{Electrons}    & Kramers   &            \\
          & Min\footnote{\textit{m} and \textit{q} are variables that control the electron excitation process attributed to the selective GAS}           & Max                & pairs     &            \\
\hline
1         &  8-\textit{m} &  8  & 4       & 6\textit{s}, 6\textit{p} \\
2         & 24-\textit{q} & 24  & 8       & 5\textit{f}, He 1\textit{s} \\
3         & 25            & 27  & 9       & 7\textit{s}, 6\textit{d}, 7\textit{p} \\
4         & 27            & 27  & $\leq$ 30 a.u.     & Virtual \\
\hline\hline
\end{tabular}
\end{table}

We report in \autoref{table1} the GAS set-up together with the technical specifications that were important in the MRCI calculation. In total, we considered 4 GAS that were selectively chosen to activate 27 electrons within 21 semi-core and valence orbitals as well as virtual orbitals with energies below 30 atomic units, i.e. 154 virtuals for the Rf$^+$-He system. Because the total number of configuration state functions was too large, we defined the parameters \textit{m} and \textit{q} to control the electron excitation process that occurred at the semi-core level. These parameters were set to \textit{m}=2 and \textit{q}=1. \textcolor{black}{It is noteworthy that the definition of the GAS in the present work enables mainly inter-electron correlation that correspond to the valence-valence and core-valence interactions. The core-core interactions are not included for minimising the configuration interaction expansion in terms of single determinants and also computational cost, which may explain the deviation between the two models: Fock-Space Coupled Cluster (FSCC) and truncated MRCI (vide infra).} Let us also note that truncated configuration interaction method is not size-consistent \cite{Szalay2012}. We did not explicitly use the Davidson (+Q) corrections \cite{DIRAC19} to solve this problem. But we surmise that including higher order excitation in the GAS scheme (see \autoref{table1}) has helped to mitigate the size-consistency issue in the present MRCI calculations. In order to validate the MRCI results, we also performed size-extensive Fock-Space Coupled Cluster (FSCC) computations with the iterative treatment of the single and double connected and disconnected excitations [\onlinecite{IH-FSCC}] in the (0,1) sector of the Fock Space. These computations used the same basis set as for the MRCI calculations, but differ for $i$) the use of the eXact-2-component Hamiltonian (X2C) \cite{X2C} to model the relativistic effects and $ii$) The omission of consideration for the basis set superposition error. We correlated all the occupied orbitals above Rf 4d and all the virtual orbitals with energy below 30 a.u. The FSCC calculations first modelled the (0,0) sector of the Fock space, which  corresponds to the closed-shell Rf$^{2+}$-He system. Subsequently, one electron was added to the (0,1) sector, that encompasses the 6d orbitals of Rf. \\
\subsection{Ion mobilities}
The ion mobilities were calculated from the ion-atom interaction potentials by solving the Boltzmann equation with the help of the Gram-Charlier approach~\cite{Viehland:2018}. To this end we used the program PC~\cite{Viehland:2010}, which delivers the momentum transfer and other transport cross sections as a function of the collision energy. From this we then calculated the reduced ion mobility $K_0$ either as a function of temperature at a given electric field-to-gas-number density (i.e., the reduced electric field, $E/n_0$) or as a function of $E/n_0$ at different gas temperatures, by means of the program VARY~\cite{Viehland:2012} and GC ~\cite{Viehland:1994}, respectively. Here $K_0$ is the ion mobility $K$ normalized to the standard pressure $P_0$ and the standard temperature $T_0$ according to $K_0=K\frac{P}{P_0}\frac{T_0}{T}$. Beyond $d=40\,$\AA, the interaction potentials were adjusted to asymptotically mimic the long-range induced ion-dipole attraction given by $V_{pol}(d)=e^2 \alpha_p/(2(4\pi\epsilon_0)^2 d^4)$, with the static average dipole polarizability of helium of $\alpha_p=0.205\,$\AA$^3$~\cite{Lide:1989}. 
The averaged mobility $K_0^{av}$~\cite{Visentin:2020} was derived based on the mobility curve calculated for each electronic state. This quantity was obtained as the average of the state-specific ion mobilities for each projection of the ion total angular momentum $J$ along the interaction axis, i.e. $\Omega$, with their statistical weights. For further insights into the ion mobility calculations, we showed the ion mobilities associated with each $\Omega$ projection of the ground $^2$D$_{3/2}$ (7s$^2$6d$^1$) state, the low-lying $^2$D$_{5/2}$ (7s$^2$6d$^1$) state and the metastable $^4$F$_{3/2}$ (7s$^1$6d$^2$) state of the Rf$^+$-He system as a Supporting Material attached to this work. \par

\begin{table}[t]
\caption{Calculated equilibrium distances $d$$_{min}$ (in {\AA}) and dissociation energies (in cm$^{-1}$) derived from the MRCI and FSCC interaction potentials for the Rf$^+$-He electronic ground state, $^2$D$_{3/2}$ (6d$^1$7s$^2$), the low-lying excited state, $^2$D$_{5/2}$ (6d$^1$7s$^2$) and the metastable state, $^4$F$_{3/2}$ (6d$^2$7s$^1$), with their $\Omega$ projections.}
\centering
\label{table2}
\begin{tabular}{l c c c c c}
\hline \hline
                               &          & \multicolumn{2}{c}{MRCI\footnote{These parameters were derived from the counterpoise energy corrected interaction potential (see in the Supporting Information Table S8 for the non-corrected values).}}   & \multicolumn{2}{c}{FSCC} \\
$^{(2S+1)}$L$_J$               &$\Omega$  & $d$$_{min}$ & $D$$_e$      &$d$$_{min}$ & $D$$_e$ \\ 
\hline 
\multirow{2}{*}{$^2$D$_{3/2}$} & 1/2      & 4.101       & 43.817       & 4.005      &59.018\\
                               & 3/2      & 4.065       & 47.875       & 3.989      &63.322\\ 
\multirow{3}{*}{$^2$D$_{5/2}$} & 1/2      & 4.193       & 39.561       & 4.095      &55.917\\
                               & 3/2      & 3.942       & 52.794       & 3.861      &68.680\\
                               & 5/2      & 4.065       & 48.503       & 4.005      &59.256\\ 
\multirow{2}{*}{$^4$F$_{3/2}$} & 1/2      & 3.987       & 46.076       &            & \\ 
                               & 3/2      & 3.864       & 53.967       &            & \\ 
\hline 
\hline
\end{tabular}
\end{table}

\section{Results and Discussion}
\subsection{Optical pumping scheme}
\textcolor{black}{Based on studies presented in Ref.\cite{Laatiaoui:2020b}} we developed a rate equation model for a four-level system to evaluate optical pumping in the Rf$^+$ ion prior to electronic state chromatography. The system consists of the ground-state $\ket{1}$: $^2$D$_{3/2}$ (7s$^2$6d$^1$), the intermediate state $\ket{2}$: $^4$F$_{3/2}$ (7s$^1$6d$^1$7p$^1$), the metastable state $\ket{3}$: $^4$F$_{3/2}$ (7s$^1$6d$^2$), and the low-lying state $\ket{4}$: $^2$D$_{5/2}$ (7s$^2$6d$^1$). \textcolor{black}{As significant energy separation between states and only higher lying odd-parity states are predicted \cite{Ramanantoanina:2021},collision-induced quenching
effects or further decay paths can be neglected}. In addition, it is assumed that broadband laser radiation is used during the initial level search such that the coherence terms in the optical Bloch equations can be safely neglected. We obtain:
\begin{eqnarray}
\nonumber\dfrac{d\rho_1}{dt}&=&\dfrac{1}{2}A_{21}S(\omega_L,\omega_{12})O(t)(\rho_2-\rho_1)\\
\nonumber&+&A_{21}\rho_2+(A_{31}^e+A_{31}^m)\rho_3+(A_{41}^e+A_{41}^m)\rho_4\\
\nonumber\dfrac{d\rho_2}{dt}&=&\dfrac{1}{2}A_{21}S(\omega_L,\omega_{12})O(t)(\rho_1-\rho_2)\\
\nonumber&-&(A_{21}+A_{23}+A_{24})\rho_2\\
\nonumber\dfrac{d\rho_3}{dt}&=&A_{23}\rho_2-(A_{31}^e+A_{31}^m+A_{34}^e+A_{34}^m)\rho_3\\
\nonumber\dfrac{d\rho_4}{dt}&=&A_{24}\rho_2+(A_{34}^e+A_{34}^m)\rho_3-(A_{41}^e+A_{41}^m)\rho_4
%\label{eq4}
%\end{dmath}
\end{eqnarray}
with the normalization $\displaystyle\sum\limits_{i}\rho_i=1$ and the initial conditions $\rho_1(t=0)=1$ and $\rho_i(t=0)=0$ for $i > 1$, where $\rho_i$, with $i=$ 1, 2, 3, 4, correspond to the occupations of individual states $\ket{i}$. $A_{ki}$, $A_{ki}^e$ and $A_{ki}^m$ are the Einstein coefficients for spontaneous emission from $\ket{k}$ to $\ket{i}$ via electric dipole ($E_1$), electric quadrupole ($E_2$) and magnetic dipole ($M_1$) transitions respectively. The values for the coefficients were taken from the work of Ramanantoanina $et$ $al.$ ~\cite{Ramanantoanina:2021}.\par
We used the frequency-dependent saturation parameter $S(\omega_L,\omega_{12})$ as described in Ref. \cite{Chhetri:2017}. 
In this parameter, we considered Doppler broadening at room temperature in terms of full width at half maximum of $0.7\,$GHz and the spectral bandwidth of the laser of $2.5\,$GHz, as well as dephasing effects on the order of $1.4\,$GHz from mode fluctuations within the laser pulse. An energy density of the laser radiation of $2.5\,\mu J/cm^2$ was taken. 
In addition, we neglected hyperfine structures and nuclear isomerism as these should be covered by the broadband laser radiation.
Moreover, the model includes a rectangular function $O(t)$ that accounts for laser radiation exposures of $10\,$ns duration only once every $100\,\mu$s period.

\begin{figure}
    \centering
    \includegraphics[width=\linewidth]{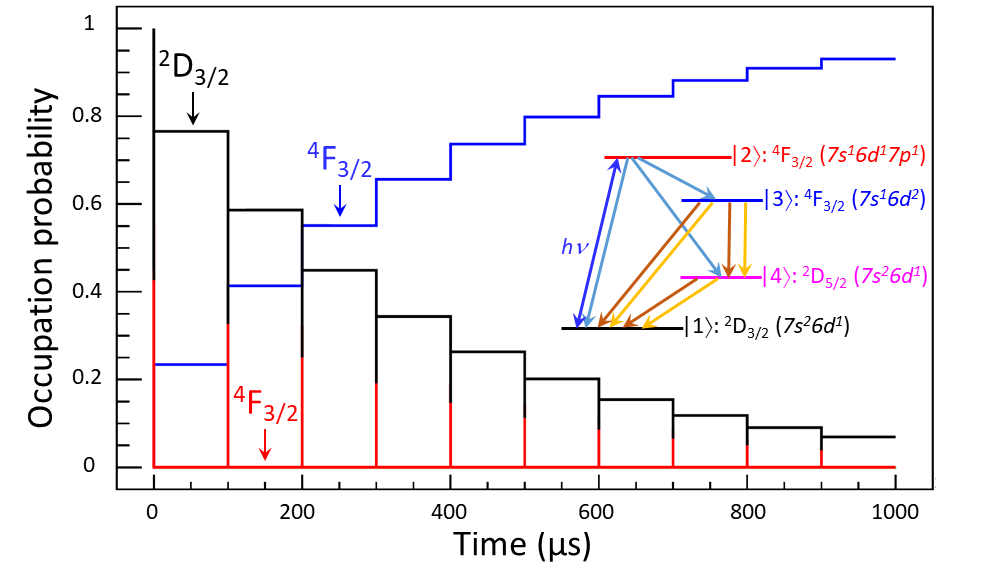}
    \caption{Laser-induced population transfer from ground- ($\ket{1}$) to metastable state ($\ket{3}$) in Rf$^+$. Level occupation is indicated for each of the modeled states in the course of $10$ laser beam exposures. The lowest-lying state ($\ket{4}$) has only a vanishing occupation probability and is not plotted in the figure. Inset: Corresponding 4-level system used in the rate-equation model with arrows in blue, yellow, and brown indicating E1, E2, and M1 transitions, respectively. Laser probing ($h\nu$) of the intermediate state ($\ket{2}:\,^4$F$_{3/2}$) induces optical pumping in the system. }
    \label{fig:pump}
\end{figure}

Figure~\ref{fig:pump} shows the calculated population of the Rf$^+$ levels in the course of $10$ laser pulses. The efficiency for optical pumping from the ground state into the metastable $^4$F$_{3/2}$ (7s$^1$6d$^2$) state saturates above $93$\% already for $10$ laser pulses, which should enable an efficient laser chromatography of this ionic species.
\subsection{Ion-atom interaction potentials}
In \autoref{table2} we list the equilibrium distances ($d$$_{min}$) and dissociation energies ($D$$_{e}$) for the Rf$^{+}$-He interaction potentials associated with each $\Omega$ projection of the ground state, $^2$D$_{3/2}$ (7s$^2$6d$^1$), the low-lying excited state, $^2$D$_{5/2}$ (7s$^2$6d$^1$), and the metastable state, $^4$F$_{3/2}$ (7s$^1$6d$^2$). A graphical representation of the interaction potentials in the minimum region is also provided in \autoref{figure2}. In addition, we also reported in the Supplementary Materials the tabulated interaction potential values at each inter-atomic distance (see Tables S1 to S7), alongside the related $d$$_{min}$ and $D$$_{e}$ values uncorrected for the basis set superposition error (see Table S8).
For the doublet states, we assess the accuracy of the MRCI results with respect to the FSCC analogs, whereas no similar assessment can be performed for the metastable $^4$F$_{3/2}$ state, due to the intrinsic limits of FSCC in the modelling of systems with more than 2 valence electrons. \\
\\
\textcolor{black}{On examination of the doublet
states}, the MRCI and FSCC methods predict a qualitatively similar behavior for the ion-atom interaction. The interaction potentials associated with the $\Omega$= 1/2 and 3/2 projections of the Rf$^+$ ground state show negligible differences and can be easily described by the average isotropic potential depicted in \autoref{figure2} (panel (b)). In contrast, the interaction potentials for  $\Omega$= 1/2, 3/2 and 5/2 of the $^2$D$_{5/2}$ (7s$^2$6d$^1$) state significantly differ, with $\Omega$= 3/2 giving rise to the most attractive interaction. Overall, both approaches model similar equilibrium distances and dissociation energies for the interaction potentials associated with the $^2$D$_{3/2}$ (7s$^2$6d$^1$) and $^2$D$_{5/2}$ (7s$^2$6d$^1$) states. \\
The two \textcolor{black}{theoretical approaches} disagree on the depth of the interaction potential. FSCC describes a more attractive and deep potential well, as shown by the smaller FSCC values for the equilibrium distances and the larger dissociation energies compared to the MRCI counterparts. In particular, for $\Omega$ = 1/2 and 3/2 the FSCC calculations predict the related potential wells to be deeper than the MRCI analogs by 19.505 and 11.143 cm$^{-1}$, respectively. We ascribe these differences to the different modelling of dynamic electron correlation provided by these two methods. We do not expect significant deviations due to \textcolor{black}{the lacking counterpoise correction} in the FSCC calculations, since the basis set superposition error accounts for no more than 1.2 cm$^{-1}$ (see Table S8 in the Supporting Materials). Remarkable discrepancies due to to the different relativistic Hamiltonians employed in the two sets of calculations (four-component Dirac-Coulomb for the MRCI computations; two-component X2C for the FSCC computations) should be excluded as well; X2C calculations usually feature accuracies comparable to those of the lengthier four-component Dirac-Coulomb counterparts even for the heaviest elements ~\cite{Dergachev:2023}. \\ 
\par
Interaction in the metastable $^4$F$_{3/2}$ (7s$^1$6d$^2$) state retains several characteristics in common with the ground and low-lying states analogs; for $\Omega$ = 3/2 the ion-atom potential well is about 17\% deeper compared to the $\Omega$ = 1/2 counterpart. However, ion-atom interaction in the quartet state features higher dissociation energies compared to interactions in the doublet states, as apparent from \autoref{figure2}. In general, the interaction is isotropic only for the ground state of the ion, whereas the doublet and quartet excited states give rise to anisotropic interaction potentials. This behavior justifies the averaging of the state-specific ion mobilities over the related $\Omega$ projections, in accordance with the so-called $anisotropic$ $approximation$ \cite{Visentin:2020, Buchachenko:2022}, rather than the averaging of the interaction potentials prior to the ion-mobility calculations.

 \begin{figure*}
\includegraphics[width=1.00\textwidth]{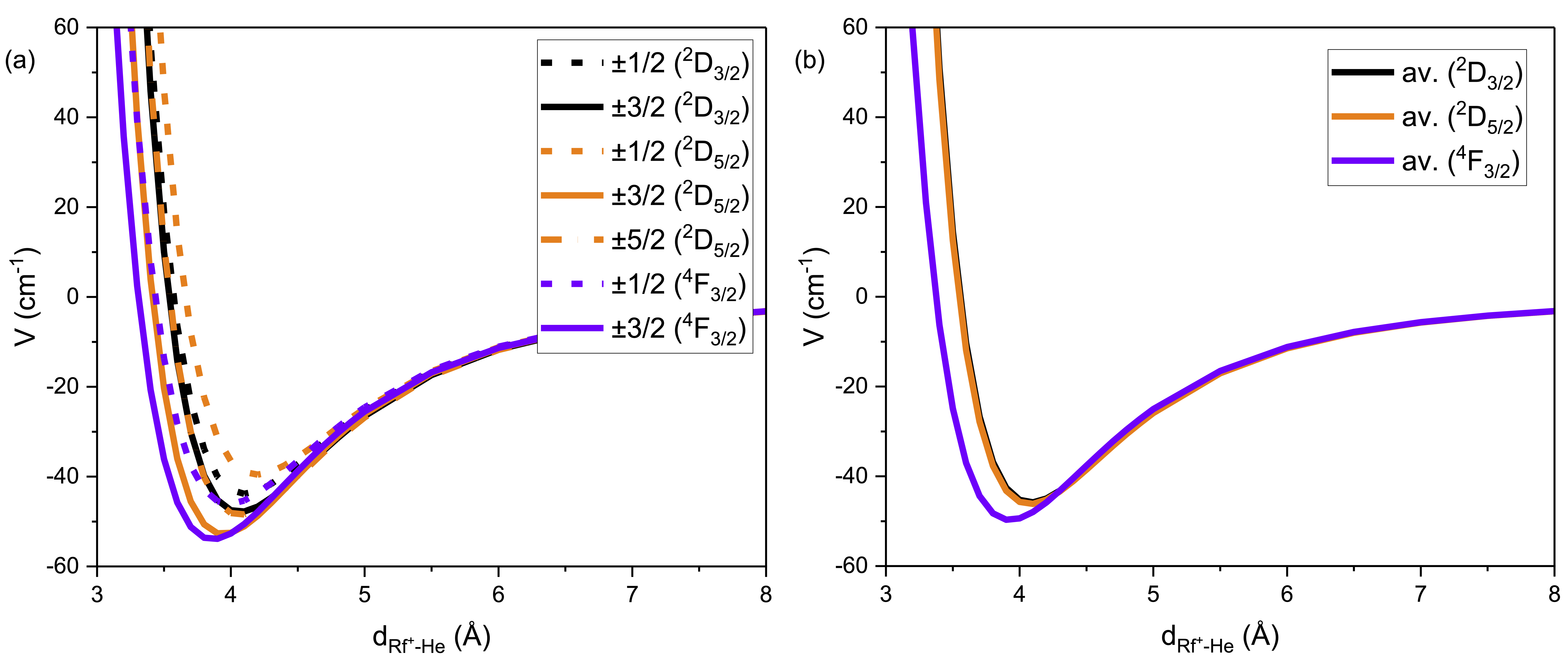}
\caption{Graphical representations of the MRCI interaction potentials of the Rf$^+$-He system  calculated for the ground state $^2$D$_{3/2}$ (7s$^2$6d$^1$), the low-lying state, $^2$D$_{5/2}$ (7s$^2$6d$^1$) and the metastable state, $^4$F$_{3/2}$ (7s$^1$6d$^2$) \textcolor{black}{for each $\Omega$ projection of the ion's total electronic angular momentum along the interaction axis} (panel (a)). Note that counterpoise energy corrections are also included in the interaction potential (see also in the Supporting Information Table S1-S7). The calculated average interaction potential for the ground, low-lying and metastable states are shown in panel (b). Note that for clarity the potentials are normalized to the same dissociation limit.}
\label{figure2}
\end{figure*}

\subsection{Ion mobility}
In order to reach significant time resolution in prospective LRC experiments on superheavy ions, we need to maximize the ESC effect and, therefore, the relative state-dependent drift time differences \cite{Laatiaoui:2020b}. This procedure depends on two experimental variables: the gas temperature $T$ and the reduced electric field $E$/$n_0$. \\
\par
\begin{figure*}
\includegraphics[width=1.00\textwidth]{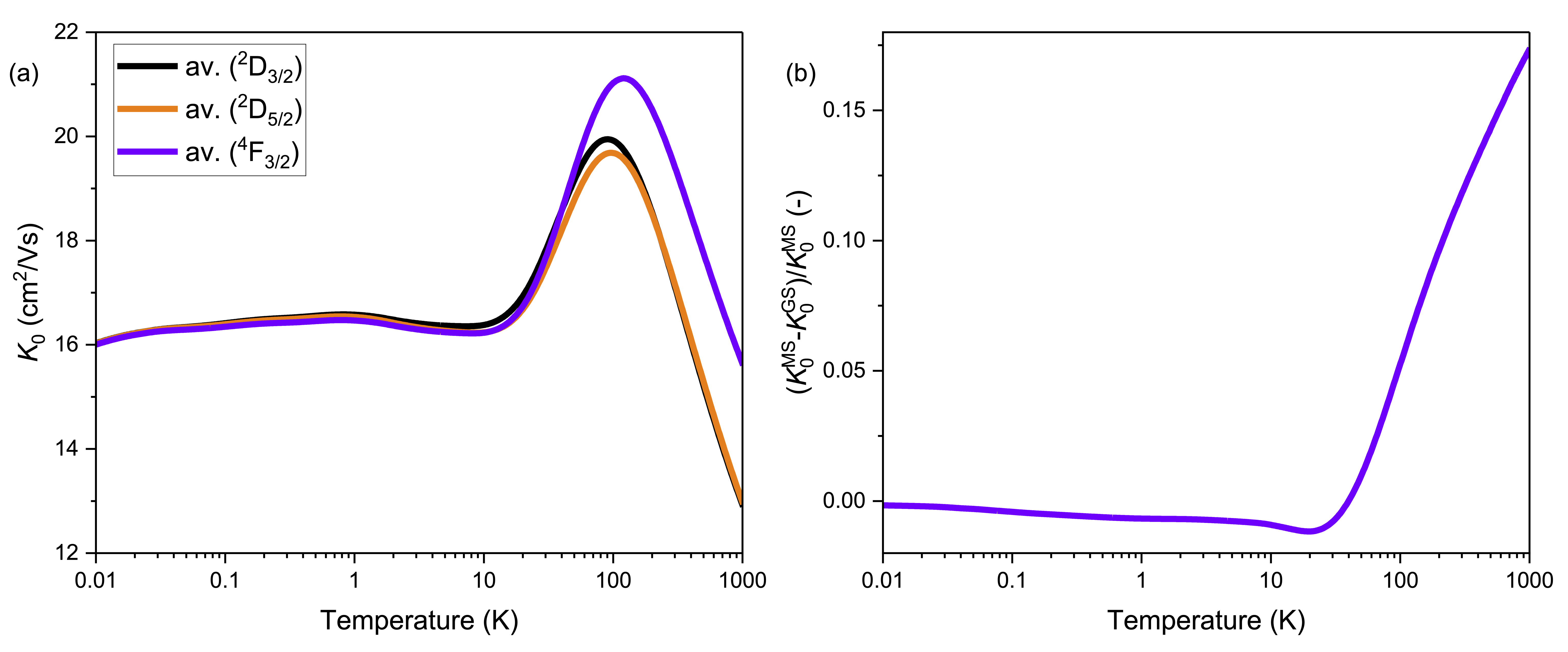}
\caption{(a) Reduced zero-field mobilities of the Rf$^+$-He system in the ground $^2$D$_{3/2}$ (7s$^2$6d$^1$) (in black), as well as in the low-lying excited $^2$D$_{5/2}$ (7s$^2$6d$^1$) (in orange) and  metastable $^4$F$_{3/2}$ (7s$^1$6d$^2$) (in violet) states as a function of the temperature, derived from the MRCI interaction potential. Note that the represented traces are average ion-mobility for the different components of the Rf$^+$-He electronic states (see also in the Supporting Information Figures S1, S2 and S3). (b) Relative differences between the reduced zero-field mobilities of the ground and the metastable states as a function of temperature.}
\label{figure3}
\end{figure*}

Let us first start from the evaluation of the reduced ion mobilities in terms of $T$. \autoref{figure3} consists of two panels: in panel (a) the averaged zero-field reduced ion mobilities of the ground $^2$D$_{3/2}$ (7s$^2$6d$^1$), low-lying $^2$D$_{5/2}$ (7s$^2$6d$^1$) and metastable $^4$F$_{3/2}$ (7s$^1$6d$^2$) states are plotted across a wide range of gas temperatures, whereas in panel (b) we depict the relative difference in the averaged zero-field ion mobilities of the metastable and the ground states. In addition, we show the reduced ion mobilities associated with each $\Omega$ projection of these states as a function of $T$ in the Figures S1, S2 and S3 of the Supplementary Material. \\With regards to panel (a), the state-dependent ion mobilities converge at the polarization limit, $K$$_{pol}$ = (13.876/$\alpha$$_p$$^{2}$) [(M$_{He}$ + M$_{ion}$)/M$_{He}$M$_{ion}$]$^{1/2}$, at 16 cm$^2$/Vs, in the mK temperature regime \cite{Viehland:1975}. Here, 13.876 is obtained when $\alpha$$_p$ is given in units of $\AA$$^3$ and the masses $M$ are in atomic mass units. The ion mobilities start to differ from around 10 K.  Higher temperatures feature a steady increase in the ion mobilities, that reach their maximum with a peak at about 100 K, followed by a monotonic decrease up to 1000 K. In this high-temperature region, the  metastable  $^4$F$_{3/2}$ state relates to a considerably larger average zero-field ion mobility compared to the ground $^2$D$_{3/2}$ and low-lying $^2$D$_{5/2}$ states. In contrast, the mobilities associated with these two latter states show a negligible difference at the peak temperature, while they overlap at higher temperatures. At 1000 K, the ion mobility for the metastable state decreases to about 15 cm$^2$/Vs, whereas the ion mobilities for the ground and low-lying excited states \textcolor{black}{further decrease} to about 13 cm$^2$/Vs. Panel (b) reflects the trends seen in panel (a). In particular, the difference in the ion mobilities for the metastable and the ground states becomes negative at around 10 K: there, the ground state is slightly more mobile in He compared to the metastable state. Upon increasing $T$, the situation is reversed and the ion mobility of the metastable state grows faster compared to the ground state analog. This behavior reflects the stronger interaction with He that affects the metastable state of Rf$^+$ compared to its ground state. The difference in the ion mobilities consequently rises and attains a maximum at about 1000 K. Therefore, the temperature range between roughly 100 and 1000 K is an optimal temperature window to detect the electronic state chromatographic effect.    \\
\par
\begin{figure*}
\includegraphics[width=1.00\textwidth]{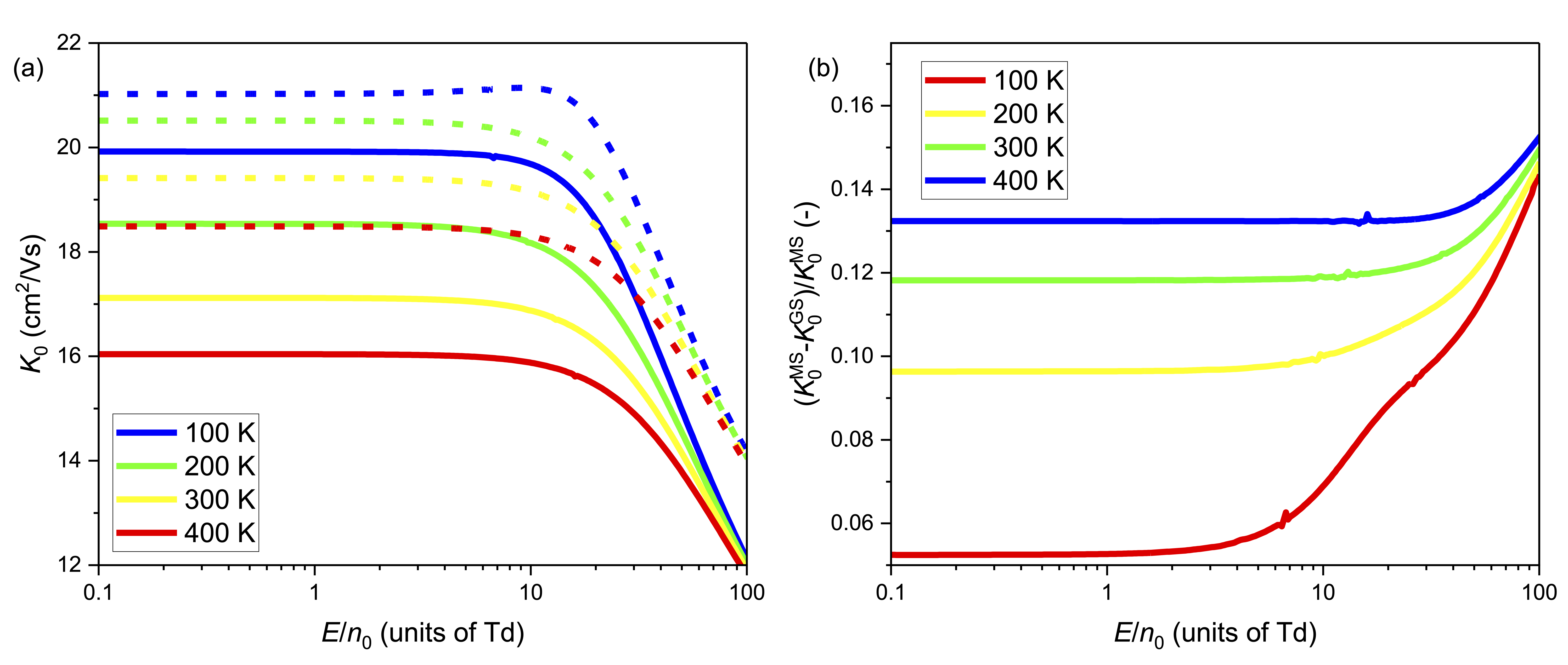}
\caption{(a) Reduced mobilities of the Rf$^+$-He  system as function of E/n$_0$ and at selected temperatures, derived from the MRCI interaction potential, corresponding to the ground $^2$D$_{3/2}$ (7s$^2$6d$^1$) (solid lines) and the metastable $^4$F$_{3/2}$ (7s$^1$6d$^2$) (dashed lines) states. The selected temperatures decrease from 400 K (lowest-lying full and dashed red lines) down to 100 K (highest-lying full and dashed blue lines).
Note that the depicted mobility curves correspond to the average calculated mobility for the state specific omega values (see also Figures S4 and S5 in the Supplementary Material). (b) Relative differences between the reduced mobilities for the ground and the metastable states of the Rf$^+$-He  system as function of E/n$_0$ at selected temperatures, derived from the MRCI interaction potentials. The operating temperatures span from 400 K (highest-lying blue line) down to 100 K (lowest-lying red line).}
\label{figure4}
\end{figure*}

We can now cast our attention to the ion mobility variation as a function of the reduced electric field. We make this comparison in \autoref{figure4}. The figure consists of two panels; in panel (a) we show the average reduced ion mobilities associated with the ground $^2$D$_{3/2}$ state (full lines) and metastable $^4$F$_{3/2}$ (dashed lines) state  at four different temperatures (100, 200, 300 and 400 K) as a function of the reduced electric field; in panel (b) the same plot is shown for the difference between the ion mobilities of the ground and the metastable states. The reduced electric field is given in units of Townsend (1 Td = \textcolor{black}{10$^{-17}$ Vcm$^{2}$}). For each state, we also showed the $\Omega$-dependent ion mobilities in terms of reduced electric field in the Figures S4, S5 and S6 of the Supplementary Materials. \\
\textcolor{black}{In Figure 4 panel (a)}, we distinguish two general features: first, the ion mobility is larger at 100 K and decreases as the gas temperature increases, in agreement with \autoref{figure3}; second, the ion mobility is approximately constant for values of the reduced electric field below 10 Td. At higher values of $E$/$n_0$ the ion mobility decreases almost exponentially, until it falls below 12 and 14 Td for the ions in the ground and metastable states, respectively. For very high values of $E$/$n_0$ it nearly decouples from the temperature dependency. This behavior was observed also for the mobilities of Lu$^+$ and Lr$^+$ in He \cite{Ramanantoanina:2023} and was ascribed to the energy gained from the electric field that dominates over the effective ion temperature.  \\
\textcolor{black}{In Figure 4 panel (b)}, the relative difference between the ground and metastable states mobilities increases with the gas temperature. Upon raising the reduced electric field, this difference grows regardless of the temperature, albeit lower $T$ ensure a higher sensitivity to reduced-field variations. The electronic state chromatography effect is largest at 100 Td, where the ion-mobility difference lies roughly between 13\% (100 K) and 14\% (400 K). At gas temperatures above 100 K, the ESC effect is already significant for reduced electric fields smaller than 10 Td, with the relative difference between the metastable and ground states that roughly ranges from 9\% (T = 200 K) up to 13\% (T = 400 K). In particular, at room temperature the ion-mobility difference lies far above 11\%. \par

\section{Summary and Conclusion}
We envisaged a strategy for an LRC experiment on Rf$^+$ in a helium buffer gas.
As the first step, we devised a rate equation model that simulates the optical pumping prior to the electronic state chromatography. The model involves the ground state of Rf$^+$, $^2$D$_{3/2}$ (7s$^2$6d$^1$) as well as three excited states, i.e., the low-lying $^2$D$_{5/2}$ (7s$^2$6d$^1$) state, the metastable $^4$F$_{3/2}$ (7s$^1$6d$^2$) state, into which the ion is optically pumped, and the intermediate $^4$F$_{3/2}$ (7s$^1$6d$^1$7p$^1$) state. Our model predicts a 93\% pumping efficiency already for 10 laser pulses, thus ensuring efficient laser resonance chromatography of the ion. 
In the second step, we investigated the interaction of Rf$^+$ with He for the ground, low-lying and metastable states of the ion, by means of accurate relativistic MRCI and FSCC calculations. The two levels of theory are in qualitative agreement and describe the interaction to be weak and isotropic for the ground state of the ion, while the excited states feature stronger and anisotropic interactions, in particular for the metastable state. \\
We then used the computed interaction potentials to model the related ion mobilities in terms of the gas temperature and the reduced electric field, and to evaluate the experimental conditions that optimize the ESC effect. For gas temperatures above 100 K and reduced fields below 100 Td, the relative drift time difference between the ground and the metastable state should lie well above 9\% and grow up to 14\% as the temperature increases. These features are indeed encouraging for prospective LRC experiments on Rf$^+$, as they show that sophisticated cryogenic ion mobility spectrometers can be avoided. Reduced fields below 10 Td and room temperatures are recommended in order to avoid the quenching of the states that occurs at high effective ion temperatures \cite{Laatiaoui:2020b} and to maintain state populations in the prospective LRC experiments \cite{RomeroRomero:2022}. In these conditions, the experiment still ensures optimal ESC effects, as the relative drift time difference between the ground and excited state should lie above 11\%, \textcolor{black}{in line with the 15\% and 13\% values previously found for Lu$^+$ and Lr$^+$ at the same temperature and reduced field conditions} (see \cite{Ramanantoanina:2023}).

\section{Acknowledgements}
This project has received funding from the European Research Council (ERC) under the European Union’s Horizon 2020 Research and Innovation Programme (Grant Agreement No. 819957). We also gratefully acknowledge high performance computing (HPC) support, time and infrastructure from the SURFsara HPC at Snellius via the HPC-Europa3 programm, the Center for Information Technology of the University of Groningen (Peregrine), the Johannes Gutenberg University of Mainz (Mogon), and the Baden-Württemberg BWUniCluster. G.V. is grateful to his former PhD supervisor, the late Prof. Alexei A. Buchachenko. To him, his suggestions and teachings the author would like to express his everlasting gratitude.

\bibliography{main}
\end{document}

\maketitle

\section{Supplementary Materials}
\beginsupplement

\begin{table}[htp]
\centering
\caption{Calculated MRCI interaction potential E$_{Rf^+-He}$, E$_{Rf^+}$ and E$_{He}$ (in atomic units) of the Rf$^+$-He, Rf$^+$-Gh and Gh-He systems, respectively, as function of the interatomic distance d (in atomic unit) obtained for the ground state $^2$D$_{3/2}$ (7s$^2$6d$^1$), $\Omega$ = 1/2, together with the corrected potential $V$ (in atomic unit). Gh represents a ghost atom without charge but carrying the the full basis sets of the He or Rf element for the counterpoise energy correction.}
\label{table1}
\resizebox{\textwidth}{!}{
\begin{tabular}{ l  c  c  c c}
\hline\hline
      d &                  E$_{Rf^+-He}$ &          E$_{Rf^+}$ &                 E$_{He}$ &               $V$ corrected\\
\hline
      3.7794522492515403& -38672.1920219849998830& -38669.3613904946978437&     -2.8617634208099165&      0.0311319305110374\\
      3.9684248617141176& -38672.1985748028018861& -38669.3613240862032399&     -2.8617633903946817&      0.0245126737936516\\
      4.1573974741766948& -38672.2036849025971605& -38669.3612717825017171&     -2.8617633625674253&      0.0193502424735925\\
      4.3463700866392712& -38672.2077810962000513& -38669.3612304810012574&     -2.8617633464309575&      0.0152127312321682\\
      4.5353426991018484& -38672.2111077524023131& -38669.3611983009977848&     -2.8617633384086605&      0.0118538870010525\\
      4.7243153115644256& -38672.2138117991999025& -38669.3611734572987189&     -2.8617633277880126&      0.0091249858887750\\
      4.9132879240270029& -38672.2159933423972689& -38669.3611540072015487&     -2.8617633116251135&      0.0069239764270606\\
      5.1022605364895801& -38672.2177319975016871& -38669.3611381771988817&     -2.8617632926781709&      0.0051694723733817\\
      5.2912331489521565& -38672.2190980848972686& -38669.3611246797008789&     -2.8617632722363107&      0.0037898670416325\\
      5.4802057614147337& -38672.2201556950967642& -38669.3611127858966938&     -2.8617632500548273&      0.0027203408535570\\
      5.6691783738773109& -38672.2209625242030597& -38669.3611021932010772&     -2.8617632287870181&      0.0019028977840208\\
      5.8581509863398882& -38672.2215690833982080& -38669.3610928281996166&     -2.8617632120003287&      0.0012869568017777\\
      6.0471235988024654& -38672.2220183062017895& -38669.3610847083982662&     -2.8617632005562852&      0.0008296027517645\\
      6.2360962112650418& -38672.2223457526997663& -38669.3610778462971211&     -2.8617631911040649&      0.0004952847011737\\
      6.4250688237276190& -38672.2225802148022922& -38669.3610721954973997&     -2.8617631796995013&      0.0002551603975007\\
      6.6140414361901954& -38672.2227445773023646& -38669.3610676423995756&     -2.8617631642818164&      0.0000862293818500\\
      6.8030140486527726& -38672.2228567468991969& -38669.3610640370970941&     -2.8617631460244111&     -0.0000295637801173\\
      6.9919866611153498& -38672.2229305560031207& -38669.3610612133998075&     -2.8617631270118795&     -0.0001062155934051\\
      7.1809592735779262& -38672.2229765673982911& -38669.3610590263997437&     -2.8617631085367603&     -0.0001544324622955\\
      7.3699318860405034& -38672.2230027499026619& -38669.3610573491023388&     -2.8617630907357805&     -0.0001823100610636\\
      7.5589044985030807& -38672.2230150585019146& -38669.3610560813031043&     -2.8617630726482073&     -0.0001959045475814\\
      7.7478771109656570& -38672.2230178718964453& -38669.3610551370002213&     -2.8617630536908774&     -0.0001996812061407\\
      7.9368497234282351& -38672.2230143889973988& -38669.3610544495022623&     -2.8617630339319384&     -0.0001969055665541\\
      8.1258223358908115& -38672.2230068870994728& -38669.3610539496003184&     -2.8617630140373813&     -0.0001899234630400\\
      8.3147949483533896& -38672.2229969887994230& -38669.3610535842017271&     -2.8617629947408458&     -0.0001804098574212\\
      8.5037675608159660& -38672.2229858135033282& -38669.3610533059982117&     -2.8617629766384436&     -0.0001695308674243\\
      8.6927401732785423& -38672.2229741307965014& -38669.3610530817968538&     -2.8617629594517791&     -0.0001580895477673\\
      8.8817127857411204& -38672.2229624543979298& -38669.3610528896970209&     -2.8617629426016649&     -0.0001466221001465\\
      9.0706853982036968& -38672.2229511206969619& -38669.3610527119017206&     -2.8617629253329113&     -0.0001354834603262\\
      9.2596580106662749& -38672.2229403376986738& -38669.3610525492986199&     -2.8617629071461779&     -0.0001248812550330\\
      9.4486306231288513& -38672.2229302241976256& -38669.3610523976967670&     -2.8617628877633070&     -0.0001149387389887\\
      9.6376032355914276& -38672.2229208335993462& -38669.3610522547023720&     -2.8617628672590496&     -0.0001057116387528\\
      9.8265758480540057& -38672.2229121810014476& -38669.3610521230002632&     -2.8617628460341935&     -0.0000972119669314\\
     10.0155484605165821& -38672.2229042485996615& -38669.3610520037982496&     -2.8617628246670970&     -0.0000894201366464\\
     10.2045210729791602& -38672.2228970032010693& -38669.3610518931018305&     -2.8617628037739462&     -0.0000823063237476\\
     10.3934936854417366& -38672.2228904053990846& -38669.3610517967972555&     -2.8617627837647297&     -0.0000758248352213\\
     10.5824662979043129& -38672.2228843980992679& -38669.3610517065972090&     -2.8617627650885922&     -0.0000699264128343\\
     10.7714389103668911& -38672.2228789380969829& -38669.3610516268963693&     -2.8617627478073784&     -0.0000645633917884\\
     10.9604115228294674& -38672.2228739744969062& -38669.3610515542022767&     -2.8617627320573749&     -0.0000596882382524\\
     11.1493841352920455& -38672.2228694639998139& -38669.3610514914980740&     -2.8617627177132672&     -0.0000552547862753\\
     11.3383567477546219& -38672.2228653618003591& -38669.3610514384999988&     -2.8617627046278766&     -0.0000512186743435\\
     11.5273293602171982& -38672.2228616298016277& -38669.3610513933017501&     -2.8617626926173529&     -0.0000475438791909\\
     11.7163019726797764& -38672.2228582321986323& -38669.3610513578969403&     -2.8617626814922730&     -0.0000441928059445\\
     11.9052745851423527& -38672.2228551364023588& -38669.3610513324965723&     -2.8617626711234587&     -0.0000411327855545\\
     12.0942471976049308& -38672.2228523099984159& -38669.3610513125022408&     -2.8617626613784495&     -0.0000383361184504\\
     12.2832198100675072& -38672.2228497317992151& -38669.3610513047970016&     -2.8617626521710768&     -0.0000357748285751\\
     12.4721924225300835& -38672.2228473680006573& -38669.3610512976010796&     -2.8617626434351822&     -0.0000334269643645\\
     12.6611650349926617& -38672.2228452066992759& -38669.3610512982995715&     -2.8617626351387764&     -0.0000312732590828\\
     12.8501376474552380& -38672.2228432183983386& -38669.3610513013991294&     -2.8617626272589485&     -0.0000292897384497\\
     13.0391102599178144& -38672.2228413920020103& -38669.3610513043022365&     -2.8617626197497543&     -0.0000274679478025\\
     13.2280828723803907& -38672.2228397076978581& -38669.3610513102976256&     -2.8617626126177083&     -0.0000257847859757\\
     13.4170554848429671& -38672.2228381513996283& -38669.3610513125968282&     -2.8617626057675776&     -0.0000242330352194\\
     13.6060280973055452& -38672.2228367122006603& -38669.3610513134990470&     -2.8617625992344284&     -0.0000227994678426\\
     13.7950007097681215& -38672.2228353773025447& -38669.3610513153980719&     -2.8617625929420685&     -0.0000214689644054\\
     13.9839733222306997& -38672.2228341361987987& -38669.3610513142994023&     -2.8617625868556158&     -0.0000202350420295\\
     14.1729459346932760& -38672.2228329861027305& -38669.3610513120001997&     -2.8617625809371336&     -0.0000190931677935\\
     14.3619185471558524& -38672.2228319108035066& -38669.3610513043968240&     -2.8617625751330085&     -0.0000180312708835\\
     14.5508911596184305& -38672.2228309069032548& -38669.3610512969025876&     -2.8617625694287065&     -0.0000170405692188\\
     14.7398637720810068& -38672.2228299693015288& -38669.3610512887971709&     -2.8617625637976833&     -0.0000161167045007\\
     14.9288363845435850& -38672.2228290947969072& -38669.3610512758023106&     -2.8617625581831256&     -0.0000152608117787\\
     15.1178089970061613& -38672.2228282717987895& -38669.3610512629020377&     -2.8617625526179693&     -0.0000144562800415\\
     15.3067816094687377& -38672.2228275051966193& -38669.3610512501036283&     -2.8617625470572143&     -0.0000137080351124\\
     15.4957542219313140& -38672.2228267820973997& -38669.3610512360974099&     -2.8617625415333974&     -0.0000130044645630\\
     15.6847268343938939& -38672.2228261035998003& -38669.3610512195009505&     -2.8617625360242647&     -0.0000123480713228\\
     15.8736994468564703& -38672.2228254697984084& -38669.3610512048035162&     -2.8617625305584427&     -0.0000117344388855\\
     16.0626720593190484& -38672.2228248686005827& -38669.3610511898004916&     -2.8617625251491550&     -0.0000111536501208\\
     17.0075351216319319& -38672.2228223600977799& -38669.3610511092992965&     -2.8617624993622708&     -0.0000087514345068\\
     17.9523981839448155& -38672.2228204861021368& -38669.3610510435028118&     -2.8617624771033108&     -0.0000069654925028\\
     18.8972612462577025& -38672.2228190700989217& -38669.3610509948994149&     -2.8617624594735531&     -0.0000056157223298\\
     20.7869873708834731& -38672.2228171512033441& -38669.3610509400969022&     -2.8617624372381201&     -0.0000037738718675\\
     24.5664396201350144& -38672.2228152315001353& -38669.3610509146965342&     -2.8617624228469998&     -0.0000018939535948\\
     30.2356179940123226& -38672.2228141412997502& -38669.3610509083009674&     -2.8617624206385144&     -0.0000008123606676\\
     37.7945224925154051& -38672.2228136576013640& -38669.3610509084974183&     -2.8617624205859236&     -0.0000003285167622\\
     47.2431531156442546& -38672.2228134617034812& -38669.3610509084028308&     -2.8617624205856735&     -0.0000001327134669\\
     58.5815098633988782& -38672.2228133844982949& -38669.3610509084974183&     -2.8617624205857228&     -0.0000000554136932\\
     69.9198666111534948& -38672.2228133567987243& -38669.3610509077989263&     -2.8617624205858161&     -0.0000000284126145\\
     75.5890449850308102& -38672.2228133500029799& -38669.3610509080972406&     -2.8617624205856895&     -0.0000000213185558\\
\hline\hline
\end{tabular}
}
\end{table}

\begin{table}[htp]
\centering
\caption{Calculated MRCI interaction potential E$_{Rf^+-He}$, E$_{Rf^+}$ and E$_{He}$ (in atomic units) of the Rf$^+$-He, Rf$^+$-Gh and Gh-He systems, respectively, as function of the interatomic distance d (in atomic unit) obtained for the ground state $^2$D$_{3/2}$ (7s$^2$6d$^1$), $\Omega$ = 3/2, together with the corrected potential $V$ (in atomic unit). Gh represents a ghost atom without charge but carrying the the full basis sets of the He or Rf element for the counterpoise energy correction.}
\label{table2}
\resizebox{\textwidth}{!}{
\begin{tabular}{ l  c  c  c c}
\hline\hline
      d &                  E$_{Rf^+-He}$ &          E$_{Rf^+}$ &                 E$_{He}$ &              $V$ corrected\\
\hline
      3.7794522492515403& -38672.1904788234969601& -38669.3612287646028562&     -2.8617634208099165&      0.0325133619189728\\
      3.9684248617141176& -38672.1975521892964025& -38669.3611884143974748&     -2.8617633903946817&      0.0253996154933702\\
      4.1573974741766948& -38672.2030559525010176& -38669.3611584241007222&     -2.8617633625674253&      0.0198658341687405\\
      4.3463700866392712& -38672.2074275833001593& -38669.3611359850983717&     -2.8617633464309575&      0.0154717482291744\\
      4.5353426991018484& -38672.2109338075970300& -38669.3611192871976527&     -2.8617633384086605&      0.0119488180062035\\
      4.7243153115644256& -38672.2137472410977352& -38669.3611067984020337&     -2.8617633277880126&      0.0091228850942571\\
      4.9132879240270029& -38672.2159907107998151& -38669.3610971544985659&     -2.8617633116251135&      0.0068697553215316\\
      5.1022605364895801& -38672.2177613059975556& -38669.3610893283985206&     -2.8617632926781709&      0.0050913150771521\\
      5.2912331489521565& -38672.2191416026034858& -38669.3610827002994483&     -2.8617632722363107&      0.0037043699339847\\
      5.4802057614147337& -38672.2202036665985361& -38669.3610769833976519&     -2.8617632500548273&      0.0026365668527433\\
      5.6691783738773109& -38672.2210100260999752& -38669.3610720744982245&     -2.8617632287870181&      0.0018252771842526\\
      5.8581509863398882& -38672.2216139040974667& -38669.3610679202029132&     -2.8617632120003287&      0.0012172281058156\\
      6.0471235988024654& -38672.2220596305996878& -38669.3610644696964300&     -2.8617632005562852&      0.0007680396520300\\
      6.2360962112650418& -38672.2223834171018098& -38669.3610616588994162&     -2.8617631911040649&      0.0004414329014253\\
      6.4250688237276190& -38672.2226143278967356& -38669.3610594076017151&     -2.8617631796995013&      0.0002082594073727\\
      6.6140414361901954& -38672.2227753502011183& -38669.3610576257997309&     -2.8617631642818164&      0.0000454398832517\\
      6.8030140486527726& -38672.2228844245983055& -38669.3610562300018501&     -2.8617631460244111&     -0.0000650485744700\\
      6.9919866611153498& -38672.2229553897996084& -38669.3610551407982712&     -2.8617631270118795&     -0.0001371219914290\\
      7.1809592735779262& -38672.2229988003018661& -38669.3610542987007648&     -2.8617631085367603&     -0.0001813930648495\\
      7.3699318860405034& -38672.2230226086976472& -38669.3610536510022939&     -2.8617630907357805&     -0.0002058669560938\\
      7.5589044985030807& -38672.2230327517972910& -38669.3610531591984909&     -2.8617630726482073&     -0.0002165199475712\\
      7.7478771109656570& -38672.2230335924978135& -38669.3610527901983005&     -2.8617630536908774&     -0.0002177486094297\\
      7.9368497234282351& -38672.2230283202006831& -38669.3610525241965661&     -2.8617630339319384&     -0.0002127620755346\\
      8.1258223358908115& -38672.2230192071001511& -38669.3610523349998402&     -2.8617630140373813&     -0.0002038580641965\\
      8.3147949483533896& -38672.2230078729990055& -38669.3610522058006609&     -2.8617629947408458&     -0.0001926724580699\\
      8.5037675608159660& -38672.2229954318027012& -38669.3610521177033661&     -2.8617629766384436&     -0.0001803374616429\\
      8.6927401732785423& -38672.2229826436014264& -38669.3610520563015598&     -2.8617629594517791&     -0.0001676278479863\\
      8.8817127857411204& -38672.2229700075986329& -38669.3610520099027781&     -2.8617629426016649&     -0.0001550550950924\\
      9.0706853982036968& -38672.2229578417973244& -38669.3610519642024883&     -2.8617629253329113&     -0.0001429522599210\\
      9.2596580106662749& -38672.2229463352996390& -38669.3610519197027315&     -2.8617629071461779&     -0.0001315084518865\\
      9.4486306231288513& -38672.2229355886011035& -38669.3610518705972936&     -2.8617628877633070&     -0.0001208302419400\\
      9.6376032355914276& -38672.2229256395003176& -38669.3610518130008131&     -2.8617628672590496&     -0.0001109592412831\\
      9.8265758480540057& -38672.2229164910022519& -38669.3610517498018453&     -2.8617628460341935&     -0.0001018951661536\\
     10.0155484605165821& -38672.2229081162004150& -38669.3610516842018114&     -2.8617628246670970&     -0.0000936073338380\\
     10.2045210729791602& -38672.2229004757027724& -38669.3610516147018643&     -2.8617628037739462&     -0.0000860572254169\\
     10.3934936854417366& -38672.2228935250968789& -38669.3610515503023635&     -2.8617627837647297&     -0.0000791910279077\\
     10.5824662979043129& -38672.2228872037012479& -38669.3610514859974501&     -2.8617627650885922&     -0.0000729526145733\\
     10.7714389103668911& -38672.2228814650006825& -38669.3610514285028330&     -2.8617627478073784&     -0.0000672886890243\\
     10.9604115228294674& -38672.2228762546001235& -38669.3610513758030720&     -2.8617627320573749&     -0.0000621467406745\\
     11.1493841352920455& -38672.2228715257006115& -38669.3610513317980804&     -2.8617627177132672&     -0.0000574761870666\\
     11.3383567477546219& -38672.2228672300989274& -38669.3610512964005466&     -2.8617627046278766&     -0.0000532290723640\\
     11.5273293602171982& -38672.2228633261984214& -38669.3610512673985795&     -2.8617626926173529&     -0.0000493661791552\\
     11.7163019726797764& -38672.2228597751964116& -38669.3610512465966167&     -2.8617626814922730&     -0.0000458471040474\\
     11.9052745851423527& -38672.2228565417026402& -38669.3610512337982072&     -2.8617626711234587&     -0.0000426367842010\\
     12.0942471976049308& -38672.2228535910035134& -38669.3610512239974923&     -2.8617626613784495&     -0.0000397056282964\\
     12.2832198100675072& -38672.2228508999032783& -38669.3610512242012192&     -2.8617626521710768&     -0.0000370235284208\\
     12.4721924225300835& -38672.2228484334991663& -38669.3610512228988227&     -2.8617626434351822&     -0.0000345671651303\\
     12.6611650349926617& -38672.2228461785998661& -38669.3610512278028182&     -2.8617626351387764&     -0.0000323156564264\\
     12.8501376474552380& -38672.2228441049010144& -38669.3610512336017564&     -2.8617626272589485&     -0.0000302440384985\\
     13.0391102599178144& -38672.2228422006010078& -38669.3610512380982982&     -2.8617626197497543&     -0.0000283427507384\\
     13.2280828723803907& -38672.2228404453999246& -38669.3610512450977694&     -2.8617626126177083&     -0.0000265876878984\\
     13.4170554848429671& -38672.2228388246003306& -38669.3610512481973274&     -2.8617626057675776&     -0.0000249706354225\\
     13.6060280973055452& -38672.2228373270991142& -38669.3610512497034506&     -2.8617625992344284&     -0.0000234781618929\\
     13.7950007097681215& -38672.2228359392975108& -38669.3610512524028309&     -2.8617625929420685&     -0.0000220939546125\\
     13.9839733222306997& -38672.2228346503979992& -38669.3610512523009675&     -2.8617625868556158&     -0.0000208112396649\\
     14.1729459346932760& -38672.2228334571991581& -38669.3610512512022979&     -2.8617625809371336&     -0.0000196250621229\\
     14.3619185471558524& -38672.2228323429008015& -38669.3610512450977694&     -2.8617625751330085&     -0.0000185226672329\\
     14.5508911596184305& -38672.2228313037994667& -38669.3610512391969678&     -2.8617625694287065&     -0.0000174951710505\\
     14.7398637720810068& -38672.2228303344018059& -38669.3610512332015787&     -2.8617625637976833&     -0.0000165374003700\\
     14.9288363845435850& -38672.2228294312008074& -38669.3610512222003308&     -2.8617625581831256&     -0.0000156508176588\\
     15.1178089970061613& -38672.2228285823002807& -38669.3610512115992606&     -2.8617625526179693&     -0.0000148180843098\\
     15.3067816094687377& -38672.2228277921967674& -38669.3610512012019171&     -2.8617625470572143&     -0.0000140439369716\\
     15.4957542219313140& -38672.2228270478008199& -38669.3610511896986281&     -2.8617625415333974&     -0.0000133165667648\\
     15.6847268343938939& -38672.2228263500001049& -38669.3610511755978223&     -2.8617625360242647&     -0.0000126383747556\\
     15.8736994468564703& -38672.2228256985981716& -38669.3610511634033173&     -2.8617625305584427&     -0.0000120046388474\\
     16.0626720593190484& -38672.2228250815969659& -38669.3610511508013587&     -2.8617625251491550&     -0.0000114056456368\\
     17.0075351216319319& -38672.2228225119979470& -38669.3610510819999035&     -2.8617624993622708&     -0.0000089306340669\\
     17.9523981839448155& -38672.2228205981009523& -38669.3610510255020927&     -2.8617624771033108&     -0.0000070954920375\\
     18.8972612462577025& -38672.2228191546018934& -38669.3610509834034019&     -2.8617624594735531&     -0.0000057117213146\\
     20.7869873708834731& -38672.2228172012983123& -38669.3610509354984970&     -2.8617624372381201&     -0.0000038285652408\\
     24.5664396201350144& -38672.2228152516036062& -38669.3610509139980422&     -2.8617624228469998&     -0.0000019147555577\\
     30.2356179940123226& -38672.2228141477971803& -38669.3610509083009674&     -2.8617624206385144&     -0.0000008188580978\\
     37.7945224925154051& -38672.2228136596022523& -38669.3610509084974183&     -2.8617624205859236&     -0.0000003305176506\\
     47.2431531156442546& -38672.2228134623001097& -38669.3610509084028308&     -2.8617624205856735&     -0.0000001333100954\\
     58.5815098633988782& -38672.2228133847966092& -38669.3610509084974183&     -2.8617624205857228&     -0.0000000557120075\\
     69.9198666111534948& -38672.2228133569005877& -38669.3610509077989263&     -2.8617624205858161&     -0.0000000285144779\\
     75.5890449850308102& -38672.2228133500975673& -38669.3610509080972406&     -2.8617624205856895&     -0.0000000214131433\\
\hline\hline
\end{tabular}
}
\end{table}

\begin{table}[htp]
\centering
\caption{Calculated MRCI interaction potential E$_{Rf^+-He}$, E$_{Rf^+}$ and E$_{He}$ (in atomic units) of the Rf$^+$-He, Rf$^+$-Gh and Gh-He systems, respectively, as function of the interatomic distance d (in atomic unit) obtained for the excited state $^2$D$_{5/2}$ (7s$^2$6d$^1$) , $\Omega$ = 1/2, together with the corrected potential $V$ (in atomic unit). Gh represents a ghost atom without charge but carrying the the full basis sets of the He or Rf element for the counterpoise energy correction.}
\label{table3}
\resizebox{\textwidth}{!}{
\begin{tabular}{ l  c  c  c c}
\hline\hline
      d &                  E$_{Rf^+-He}$ &          E$_{Rf^+}$ &                 E$_{He}$ &              $V$ corrected\\
\hline
      3.7794522492515403& -38672.1540908992028562& -38669.3347726927968324&     -2.8617634208099165&      0.0424452144070528\\
      3.9684248617141176& -38672.1634975912020309& -38669.3347018506028689&     -2.8617633903946817&      0.0329676497931359\\
      4.1573974741766948& -38672.1708581503989990& -38669.3346449813980144&     -2.8617633625674253&      0.0255501935680513\\
      4.3463700866392712& -38672.1766509106018930& -38669.3345992156973807&     -2.8617633464309575&      0.0197116515264497\\
      4.5353426991018484& -38672.1812174674996641& -38669.3345629391988041&     -2.8617633384086605&      0.0151088101047208\\
      4.7243153115644256& -38672.1848119392016088& -38669.3345345772977453&     -2.8617633277880126&      0.0114859658860951\\
      4.9132879240270029& -38672.1876290255022468& -38669.3345122476021061&     -2.8617633116251135&      0.0086465337226400\\
      5.1022605364895801& -38672.1898222067975439& -38669.3344940851966385&     -2.8617632926781709&      0.0064351710752817\\
      5.2912331489521565& -38672.1915155797032639& -38669.3344786305970047&     -2.8617632722363107&      0.0047263231317629\\
      5.4802057614147337& -38672.1928109509972273& -38669.3344649865030078&     -2.8617632500548273&      0.0034172855594079\\
      5.6691783738773109& -38672.1937920593991294& -38669.3344527414010372&     -2.8617632287870181&      0.0024239107879112\\
      5.8581509863398882& -38672.1945273813980748& -38669.3344417904008878&     -2.8617632120003287&      0.0016776210031821\\
      6.0471235988024654& -38672.1950723475965788& -38669.3344321783006308&     -2.8617632005562852&      0.0011230312593398\\
      6.2360962112650418& -38672.1954713240993442& -38669.3344239665020723&     -2.8617631911040649&      0.0007158335065469\\
      6.4250688237276190& -38672.1957593949991860& -38669.3344171462013037&     -2.8617631796995013&      0.0004209309045109\\
      6.6140414361901954& -38672.1959640023997054& -38669.3344116146981833&     -2.8617631642818164&      0.0002107765831170\\
      6.8030140486527726& -38672.1961063962007756& -38669.3344072078034515&     -2.8617631460244111&      0.0000639576246613\\
      6.9919866611153498& -38672.1962028908019420& -38669.3344037319984636&     -2.8617631270118795&     -0.0000360317935701\\
      7.1809592735779262& -38672.1962659092023387& -38669.3344010130967945&     -2.8617631085367603&     -0.0001017875692924\\
      7.3699318860405034& -38672.1963048250036081& -38669.3343988998021814&     -2.8617630907357805&     -0.0001428344621672\\
      7.5589044985030807& -38672.1963266562015633& -38669.3343972743023187&     -2.8617630726482073&     -0.0001663092480158\\
      7.7478771109656570& -38672.1963365847987006& -38669.3343960372003494&     -2.8617630536908774&     -0.0001774939082679\\
      7.9368497234282351& -38672.1963384148984915& -38669.3343951110000489&     -2.8617630339319384&     -0.0001802699698601\\
      8.1258223358908115& -38672.1963348775025224& -38669.3343944173029740&     -2.8617630140373813&     -0.0001774461634341\\
      8.3147949483533896& -38672.1963279354022234& -38669.3343938947000424&     -2.8617629947408458&     -0.0001710459619062\\
      8.5037675608159660& -38672.1963189625967061& -38669.3343934898002772&     -2.8617629766384436&     -0.0001624961587368\\
      8.6927401732785423& -38672.1963089206983568& -38669.3343931646013516&     -2.8617629594517791&     -0.0001527966451249\\
      8.8817127857411204& -38672.1962984674973995& -38669.3343928941030754&     -2.8617629426016649&     -0.0001426307935617\\
      9.0706853982036968& -38672.1962880477003637& -38669.3343926581001142&     -2.8617629253329113&     -0.0001324642653344\\
      9.2596580106662749& -38672.1962779501991463& -38669.3343924549990334&     -2.8617629071461779&     -0.0001225880550919\\
      9.4486306231288513& -38672.1962683535966789& -38669.3343922778003616&     -2.8617628877633070&     -0.0001131880344474\\
      9.6376032355914276& -38672.1962593542994000& -38669.3343921202977072&     -2.8617628672590496&     -0.0001043667434715\\
      9.8265758480540057& -38672.1962509985023644& -38669.3343919812032254&     -2.8617628460341935&     -0.0000961712648859\\
     10.0155484605165821& -38672.1962432906002505& -38669.3343918578975718&     -2.8617628246670970&     -0.0000886080379132\\
     10.2045210729791602& -38672.1962362133999704& -38669.3343917426973348&     -2.8617628037739462&     -0.0000816669271444\\
     10.3934936854417366& -38672.1962297392965411& -38669.3343916388985235&     -2.8617627837647297&     -0.0000753166314098\\
     10.5824662979043129& -38672.1962238207997871& -38669.3343915372970514&     -2.8617627650885922&     -0.0000695184135111\\
     10.7714389103668911& -38672.1962184221993084& -38669.3343914419965586&     -2.8617627478073784&     -0.0000642323939246\\
     10.9604115228294674& -38672.1962134993009386& -38669.3343913502976648&     -2.8617627320573749&     -0.0000594169468968\\
     11.1493841352920455& -38672.1962090145971160& -38669.3343912666023243&     -2.8617627177132672&     -0.0000550302793272\\
     11.3383567477546219& -38672.1962049282010412& -38669.3343911924021086&     -2.8617627046278766&     -0.0000510311729158\\
     11.5273293602171982& -38672.1962012059011613& -38669.3343911275005667&     -2.8617626926173529&     -0.0000473857799079\\
     11.7163019726797764& -38672.1961978152030497& -38669.3343910750991199&     -2.8617626814922730&     -0.0000440586081822\\
     11.9052745851423527& -38672.1961947254967527& -38669.3343910365001648&     -2.8617626711234587&     -0.0000410178763559\\
     12.0942471976049308& -38672.1961919056993793& -38669.3343910070980201&     -2.8617626613784495&     -0.0000382372236345\\
     12.2832198100675072& -38672.1961893351981416& -38669.3343909938994329&     -2.8617626521710768&     -0.0000356891250703\\
     12.4721924225300835& -38672.1961869805963943& -38669.3343909846007591&     -2.8617626434351822&     -0.0000333525604219\\
     12.6611650349926617& -38672.1961848294013180& -38669.3343909860996064&     -2.8617626351387764&     -0.0000312081610900\\
     12.8501376474552380& -38672.1961828518979019& -38669.3343909920004080&     -2.8617626272589485&     -0.0000292326367344\\
     13.0391102599178144& -38672.1961810365028214& -38669.3343909988980158&     -2.8617626197497543&     -0.0000274178528343\\
     13.2280828723803907& -38672.1961793629016029& -38669.3343910096009495&     -2.8617626126177083&     -0.0000257406863966\\
     13.4170554848429671& -38672.1961778166005388& -38669.3343910167022841&     -2.8617626057675776&     -0.0000241941306740\\
     13.6060280973055452& -38672.1961763864965178& -38669.3343910219991812&     -2.8617625992344284&     -0.0000227652635658\\
     13.7950007097681215& -38672.1961750595000922& -38669.3343910277035320&     -2.8617625929420685&     -0.0000214388564928\\
     13.9839733222306997& -38672.1961738249010523& -38669.3343910296025570&     -2.8617625868556158&     -0.0000202084411285\\
     14.1729459346932760& -38672.1961726801018813& -38669.3343910295006935&     -2.8617625809371336&     -0.0000190696664504\\
     14.3619185471558524& -38672.1961716087971581& -38669.3343910230978508&     -2.8617625751330085&     -0.0000180105635081\\
     14.5508911596184305& -38672.1961706078000134& -38669.3343910158000654&     -2.8617625694287065&     -0.0000170225684997\\
     14.7398637720810068& -38672.1961696720027248& -38669.3343910073017469&     -2.8617625637976833&     -0.0000161009011208\\
     14.9288363845435850& -38672.1961687981965952& -38669.3343909930990776&     -2.8617625581831256&     -0.0000152469146997\\
     15.1178089970061613& -38672.1961679751984775& -38669.3343909784016432&     -2.8617625526179693&     -0.0000144441801240\\
     15.3067816094687377& -38672.1961672078978154& -38669.3343909632021678&     -2.8617625470572143&     -0.0000136976377689\\
     15.4957542219313140& -38672.1961664835034753& -38669.3343909464965691&     -2.8617625415333974&     -0.0000129954714794\\
     15.6847268343938939& -38672.1961658032014384& -38669.3343909268005518&     -2.8617625360242647&     -0.0000123403733596\\
     15.8736994468564703& -38672.1961651671008440& -38669.3343909086979693&     -2.8617625305584427&     -0.0000117278468679\\
     16.0626720593190484& -38672.1961645635019522& -38669.3343908901006216&     -2.8617625251491550&     -0.0000111482513603\\
     17.0075351216319319& -38672.1961620407018927& -38669.3343907915987074&     -2.8617624993622708&     -0.0000087497392087\\
     17.9523981839448155& -38672.1961601544026053& -38669.3343907111993758&     -2.8617624771033108&     -0.0000069660964073\\
     18.8972612462577025& -38672.1961587303012493& -38669.3343906532973051&     -2.8617624594735531&     -0.0000056175267673\\
     20.7869873708834731& -38672.1961568058031844& -38669.3343905919973622&     -2.8617624372381201&     -0.0000037765712477\\
     24.5664396201350144& -38672.1961548862018390& -38669.3343905669025844&     -2.8617624228469998&     -0.0000018964492483\\
     30.2356179940123226& -38672.1961537957031396& -38669.3343905608999194&     -2.8617624206385144&     -0.0000008141651051\\
     37.7945224925154051& -38672.1961533111971221& -38669.3343905611982336&     -2.8617624205859236&     -0.0000003294117050\\
     47.2431531156442546& -38672.1961531147971982& -38669.3343905610963702&     -2.8617624205856735&     -0.0000001331136446\\
     58.5815098633988782& -38672.1961530374974245& -38669.3343905611982336&     -2.8617624205857228&     -0.0000000557120075\\
     69.9198666111534948& -38672.1961530096014030& -38669.3343905603978783&     -2.8617624205858161&     -0.0000000286163413\\
     75.5890449850308102& -38672.1961530027983827& -38669.3343905607980560&     -2.8617624205856895&     -0.0000000214131433\\
\hline\hline
\end{tabular}
}
\end{table}

\begin{table}[htp]
\centering
\caption{Calculated MRCI interaction potential E$_{Rf^+-He}$, E$_{Rf^+}$ and E$_{He}$ (in atomic units) of the Rf$^+$-He, Rf$^+$-Gh and Gh-He systems, respectively, as function of the interatomic distance d (in atomic unit) obtained for the excited state $^2$D$_{5/2}$ (7s$^2$6d$^1$) , $\Omega$ = 3/2, together with the corrected potential $V$ (in atomic unit). Gh represents a ghost atom without charge but carrying the the full basis sets of the He or Rf element for the counterpoise energy correction.}
\label{table4}
\resizebox{\textwidth}{!}{
\begin{tabular}{ l  c  c  c c}
\hline\hline
      d &                  E$_{Rf^+-He}$ &          E$_{Rf^+}$ &                 E$_{He}$ &              $V$ corrected\\
\hline
     3.7794522492515403& -38672.1691837640973972& -38669.3346524148000753&     -2.8617634208099165&      0.0272320715157548\\
      3.9684248617141176& -38672.1757534525968367& -38669.3346000769015518&     -2.8617633903946817&      0.0206100146970130\\
      4.1573974741766948& -38672.1806179979030276& -38669.3345595479986514&     -2.8617633625674253&      0.0157049126646598\\
      4.3463700866392712& -38672.1843188932980411& -38669.3345280217981781&     -2.8617633464309575&      0.0119724749310990\\
      4.5353426991018484& -38672.1871876568984590& -38669.3345036753016757&     -2.8617633384086605&      0.0090793568087975\\
      4.7243153115644256& -38672.1894314776000101& -38669.3344848405031371&     -2.8617633277880126&      0.0068166906930855\\
      4.9132879240270029& -38672.1911869206014671& -38669.3344698902001255&     -2.8617633116251135&      0.0050462812214391\\
      5.1022605364895801& -38672.1925518548014225& -38669.3344575024020742&     -2.8617632926781709&      0.0036689402768388\\
      5.2912331489521565& -38672.1936024966998957& -38669.3344468268987839&     -2.8617632722363107&      0.0026076024369104\\
      5.4802057614147337& -38672.1944013160027680& -38669.3344374481966952&     -2.8617632500548273&      0.0017993822475546\\
      5.6691783738773109& -38672.1950003878009738& -38669.3344292179972399&     -2.8617632287870181&      0.0011920589822694\\
      5.8581509863398882& -38672.1954429339020862& -38669.3344220833969302&     -2.8617632120003287&      0.0007423614952131\\
      6.0471235988024654& -38672.1957643529967754& -38669.3344160035994719&     -2.8617632005562852&      0.0004148511579842\\
      6.2360962112650418& -38672.1959932128011133& -38669.3344109148019925&     -2.8617631911040649&      0.0001808931046980\\
      6.4250688237276190& -38672.1961522221026826& -38669.3344067220969009&     -2.8617631796995013&      0.0000176796966116\\
      6.6140414361901954& -38672.1962591949995840& -38669.3344033086978015&     -2.8617631642818164&     -0.0000927220171434\\
      6.8030140486527726& -38672.1963279335031984& -38669.3344005601029494&     -2.8617631460244111&     -0.0001642273782636\\
      6.9919866611153498& -38672.1963690109987510& -38669.3343983653976466&     -2.8617631270118795&     -0.0002075185911963\\
      7.1809592735779262& -38672.1963904357980937& -38669.3343966355969314&     -2.8617631085367603&     -0.0002306916649104\\
      7.3699318860405034& -38672.1963981945009436& -38669.3343952897994313&     -2.8617630907357805&     -0.0002398139622528\\
      7.5589044985030807& -38672.1963967174015124& -38669.3343942623032490&     -2.8617630726482073&     -0.0002393824470346\\
      7.7478771109656570& -38672.1963892212006613& -38669.3343934923032066&     -2.8617630536908774&     -0.0002326752073714\\
      7.9368497234282351& -38672.1963780286969268& -38669.3343929332986590&     -2.8617630339319384&     -0.0002220614696853\\
      8.1258223358908115& -38672.1963647611992201& -38669.3343925340013811&     -2.8617630140373813&     -0.0002092131617246\\
      8.3147949483533896& -38672.1963505530002294& -38669.3343922557978658&     -2.8617629947408458&     -0.0001953024620889\\
      8.5037675608159660& -38672.1963361589005217& -38669.3343920624974999&     -2.8617629766384436&     -0.0001811197653296\\
      8.6927401732785423& -38672.1963220750985784& -38669.3343919264007127&     -2.8617629594517791&     -0.0001671892459854\\
      8.8817127857411204& -38672.1963086087998818& -38669.3343918268001289&     -2.8617629426016649&     -0.0001538393989904\\
      9.0706853982036968& -38672.1962959388984018& -38669.3343917431993759&     -2.8617629253329113&     -0.0001412703641108\\
      9.2596580106662749& -38672.1962841541026137& -38669.3343916715966770&     -2.8617629071461779&     -0.0001295753609156\\
      9.4486306231288513& -38672.1962732829997549& -38669.3343916035009897&     -2.8617628877633070&     -0.0001187917368952\\
      9.6376032355914276& -38672.1962633123985142& -38669.3343915329023730&     -2.8617628672590496&     -0.0001089122379199\\
      9.8265758480540057& -38672.1962542086985195& -38669.3343914608994965&     -2.8617628460341935&     -0.0000999017647700\\
     10.0155484605165821& -38672.1962459194983239& -38669.3343913892022101&     -2.8617628246670970&     -0.0000917056313483\\
     10.2045210729791602& -38672.1962383873978979& -38669.3343913153003086&     -2.8617628037739462&     -0.0000842683220981\\
     10.3934936854417366& -38672.1962315561031573& -38669.3343912474010722&     -2.8617627837647297&     -0.0000775249354774\\
     10.5824662979043129& -38672.1962253567035077& -38669.3343911798001500&     -2.8617627650885922&     -0.0000714118141332\\
     10.7714389103668911& -38672.1962197374014067& -38669.3343911189003848&     -2.8617627478073784&     -0.0000658706921968\\
     10.9604115228294674& -38672.1962146407022374& -38669.3343910624025739&     -2.8617627320573749&     -0.0000608462432865\\
     11.1493841352920455& -38672.1962100183009170& -38669.3343910143012181&     -2.8617627177132672&     -0.0000562862842344\\
     11.3383567477546219& -38672.1962058209028328& -38669.3343909743998665&     -2.8617627046278766&     -0.0000521418769495\\
     11.5273293602171982& -38672.1962020071005099& -38669.3343909409013577&     -2.8617626926173529&     -0.0000483735784655\\
     11.7163019726797764& -38672.1961985380985425& -38669.3343909155009896&     -2.8617626814922730&     -0.0000449411018053\\
     11.9052745851423527& -38672.1961953791978885& -38669.3343908983006258&     -2.8617626711234587&     -0.0000418097770307\\
     12.0942471976049308& -38672.1961924960996839& -38669.3343908843016834&     -2.8617626613784495&     -0.0000389504202758\\
     12.2832198100675072& -38672.1961898664012551& -38669.3343908808019478&     -2.8617626521710768&     -0.0000363334256690\\
     12.4721924225300835& -38672.1961874557018746& -38669.3343908759998158&     -2.8617626434351822&     -0.0000339362668456\\
     12.6611650349926617& -38672.1961852512977202& -38669.3343908778988407&     -2.8617626351387764&     -0.0000317382582580\\
     12.8501376474552380& -38672.1961832235028851& -38669.3343908811002620&     -2.8617626272589485&     -0.0000297151418636\\
     13.0391102599178144& -38672.1961813610978425& -38669.3343908835013281&     -2.8617626197497543&     -0.0000278578445432\\
     13.2280828723803907& -38672.1961796440009493& -38669.3343908889000886&     -2.8617626126177083&     -0.0000261424866039\\
     13.4170554848429671& -38672.1961780580968480& -38669.3343908906972501&     -2.8617626057675776&     -0.0000245616320171\\
     13.6060280973055452& -38672.1961765922969789& -38669.3343908911992912&     -2.8617625992344284&     -0.0000231018639170\\
     13.7950007097681215& -38672.1961752338029328& -38669.3343908933020430&     -2.8617625929420685&     -0.0000217475608224\\
     13.9839733222306997& -38672.1961739715989097& -38669.3343908928000019&     -2.8617625868556158&     -0.0000204919415410\\
     14.1729459346932760& -38672.1961728030000813& -38669.3343908917013323&     -2.8617625809371336&     -0.0000193303640117\\
     14.3619185471558524& -38672.1961717113008490& -38669.3343908856986673&     -2.8617625751330085&     -0.0000182504663826\\
     14.5508911596184305& -38672.1961706931979279& -38669.3343908801980433&     -2.8617625694287065&     -0.0000172435684362\\
     14.7398637720810068& -38672.1961697430015192& -38669.3343908746028319&     -2.8617625637976833&     -0.0000163045988302\\
     14.9288363845435850& -38672.1961688575029257& -38669.3343908643000759&     -2.8617625581831256&     -0.0000154350200319\\
     15.1178089970061613& -38672.1961680250024074& -38669.3343908543974976&     -2.8617625526179693&     -0.0000146179881995\\
     15.3067816094687377& -38672.1961672499965061& -38669.3343908446986461&     -2.8617625470572143&     -0.0000138582399813\\
     15.4957542219313140& -38672.1961665195995010& -38669.3343908340029884&     -2.8617625415333974&     -0.0000131440610858\\
     15.6847268343938939& -38672.1961658346990589& -38669.3343908207025379&     -2.8617625360242647&     -0.0000124779689941\\
     15.8736994468564703& -38672.1961651952005923& -38669.3343908093011123&     -2.8617625305584427&     -0.0000118553434731\\
     16.0626720593190484& -38672.1961645892006345& -38669.3343907974995091&     -2.8617625251491550&     -0.0000112665511551\\
     17.0075351216319319& -38672.1961620630972902& -38669.3343907318994752&     -2.8617624993622708&     -0.0000088318338385\\
     17.9523981839448155& -38672.1961601785005769& -38669.3343906774025527&     -2.8617624771033108&     -0.0000070239912020\\
     18.8972612462577025& -38672.1961587549012620& -38669.3343906363006681&     -2.8617624594735531&     -0.0000056591234170\\
     20.7869873708834731& -38672.1961568246988463& -38669.3343905887013534&     -2.8617624372381201&     -0.0000037987629184\\
     24.5664396201350144& -38672.1961548932013102& -38669.3343905666988576&     -2.8617624228469998&     -0.0000019036524463\\
     30.2356179940123226& -38672.1961537971001235& -38669.3343905608999194&     -2.8617624206385144&     -0.0000008155620890\\
     37.7945224925154051& -38672.1961533112989855& -38669.3343905611982336&     -2.8617624205859236&     -0.0000003295135684\\
     47.2431531156442546& -38672.1961531147026108& -38669.3343905610017828&     -2.8617624205856735&     -0.0000001331136446\\
     58.5815098633988782& -38672.1961530373009737& -38669.3343905611982336&     -2.8617624205857228&     -0.0000000555155566\\
     69.9198666111534948& -38672.1961530094995396& -38669.3343905603978783&     -2.8617624205858161&     -0.0000000285144779\\
     75.5890449850308102& -38672.1961530026965193& -38669.3343905607980560&     -2.8617624205856895&     -0.0000000213112799\\
\hline\hline
\end{tabular}
}
\end{table}

\begin{table}[htp]
\centering
\caption{Calculated MRCI interaction potential E$_{Rf^+-He}$, E$_{Rf^+}$ and E$_{He}$ (in atomic units) of the Rf$^+$-He, Rf$^+$-Gh and Gh-He systems, respectively, as function of the interatomic distance d (in atomic unit) obtained for the excited state $^2$D$_{5/2}$ (7s$^2$6d$^1$) , $\Omega$ = 5/2, together with the corrected potential $V$ (in atomic unit). Gh represents a ghost atom without charge but carrying the the full basis sets of the He or Rf element for the counterpoise energy correction.}
\label{table5}
\resizebox{\textwidth}{!}{
\begin{tabular}{ l  c  c  c c}
\hline\hline
      d &                  E$_{Rf^+-He}$ &          E$_{Rf^+}$ &                 E$_{He}$ &              $V$ corrected\\
\hline
      3.7794522492515403& -38672.1605179972029873& -38669.3345455421003862&     -2.8617634208099165&      0.0357909657104756\\
      3.9684248617141176& -38672.1685201832005987& -38669.3345090353031992&     -2.8617633903946817&      0.0277522424948984\\
      4.1573974741766948& -38672.1747001421972527& -38669.3344822557992302&     -2.8617633625674253&      0.0215454761710134\\
      4.3463700866392712& -38672.1795605915976921& -38669.3344624656019732&     -2.8617633464309575&      0.0166652204352431\\
      4.5353426991018484& -38672.1834184891995392& -38669.3344479039005819&     -2.8617633384086605&      0.0127927531066234\\
      4.7243153115644256& -38672.1864840355992783& -38669.3344371174025582&     -2.8617633277880126&      0.0097164095932385\\
      4.9132879240270029& -38672.1889076950028539& -38669.3344288516018423&     -2.8617633116251135&      0.0072844682217692\\
      5.1022605364895801& -38672.1908066731994040& -38669.3344221864026622&     -2.8617632926781709&      0.0053788058794453\\
      5.2912331489521565& -38672.1922781110988581& -38669.3344165773014538&     -2.8617632722363107&      0.0039017384406179\\
      5.4802057614147337& -38672.1934045933012385& -38669.3344117746019037&     -2.8617632500548273&      0.0027704313542927\\
      5.6691783738773109& -38672.1942562340991572& -38669.3344076845969539&     -2.8617632287870181&      0.0019146792838001\\
      5.8581509863398882& -38672.1948917254994740& -38669.3344042531025480&     -2.8617632120003287&      0.0012757396034431\\
      6.0471235988024654& -38672.1953593388025183& -38669.3344014282993157&     -2.8617632005562852&      0.0008052900520852\\
      6.2360962112650418& -38672.1956981241019093& -38669.3343991504007136&     -2.8617631911040649&      0.0004642174026230\\
      6.4250688237276190& -38672.1959391861018958& -38669.3343973469964112&     -2.8617631796995013&      0.0002213405969087\\
      6.6140414361901954& -38672.1961069735989440& -38669.3343959374979022&     -2.8617631642818164&      0.0000521281835972\\
      6.8030140486527726& -38672.1962204687006306& -38669.3343948480978725&     -2.8617631460244111&     -0.0000624745807727\\
      6.9919866611153498& -38672.1962942473983276& -38669.3343940077975276&     -2.8617631270118795&     -0.0001371125908918\\
      7.1809592735779262& -38672.1963393827973050& -38669.3343933647993254&     -2.8617631085367603&     -0.0001829094617278\\
      7.3699318860405034& -38672.1963641894035391& -38669.3343928735994268&     -2.8617630907357805&     -0.0002082250648527\\
      7.5589044985030807& -38672.1963748537018546& -38669.3343925025983481&     -2.8617630726482073&     -0.0002192784522776\\
      7.7478771109656570& -38672.1963759092977853& -38669.3343922245985596&     -2.8617630536908774&     -0.0002206310091424\\
      7.9368497234282351& -38672.1963706609021756& -38669.3343920257029822&     -2.8617630339319384&     -0.0002156012706109\\
      8.1258223358908115& -38672.1963614576015971& -38669.3343918846003362&     -2.8617630140373813&     -0.0002065589651465\\
      8.3147949483533896& -38672.1963499691992183& -38669.3343917884994880&     -2.8617629947408458&     -0.0001951859594556\\
      8.5037675608159660& -38672.1963373421967844& -38669.3343917222009622&     -2.8617629766384436&     -0.0001826433581300\\
      8.6927401732785423& -38672.1963243572972715& -38669.3343916740996065&     -2.8617629594517791&     -0.0001697237457847\\
      8.8817127857411204& -38672.1963115261023631& -38669.3343916352023371&     -2.8617629426016649&     -0.0001569482992636\\
      9.0706853982036968& -38672.1962991745967884& -38669.3343915933000972&     -2.8617629253329113&     -0.0001446559617762\\
      9.2596580106662749& -38672.1962874956006999& -38669.3343915502991877&     -2.8617629071461779&     -0.0001330381564912\\
      9.4486306231288513& -38672.1962765919015510& -38669.3343915017030668&     -2.8617628877633070&     -0.0001222024366143\\
      9.6376032355914276& -38672.1962665016035317& -38669.3343914443976246&     -2.8617628672590496&     -0.0001121899476857\\
      9.8265758480540057& -38672.1962572272968828& -38669.3343913818971487&     -2.8617628460341935&     -0.0001029993654811\\
     10.0155484605165821& -38672.1962487409982714& -38669.3343913173011970&     -2.8617628246670970&     -0.0000945990323089\\
     10.2045210729791602& -38672.1962410017004004& -38669.3343912491982337&     -2.8617628037739462&     -0.0000869487266755\\
     10.3934936854417366& -38672.1962339638994308& -38669.3343911863994435&     -2.8617627837647297&     -0.0000799937333795\\
     10.5824662979043129& -38672.1962275652986136& -38669.3343911237025168&     -2.8617627650885922&     -0.0000736765068723\\
     10.7714389103668911& -38672.1962217578984564& -38669.3343910678013344&     -2.8617627478073784&     -0.0000679422882968\\
     10.9604115228294674& -38672.1962164864016813& -38669.3343910163966939&     -2.8617627320573749&     -0.0000627379486104\\
     11.1493841352920455& -38672.1962117029033834& -38669.3343909734976478&     -2.8617627177132672&     -0.0000580116902711\\
     11.3383567477546219& -38672.1962073583999882& -38669.3343909390023327&     -2.8617627046278766&     -0.0000537147716386\\
     11.5273293602171982& -38672.1962034106982173& -38669.3343909106988576&     -2.8617626926173529&     -0.0000498073786730\\
     11.7163019726797764& -38672.1961998203987605& -38669.3343908905007993&     -2.8617626814922730&     -0.0000462484022137\\
     11.9052745851423527& -38672.1961965516966302& -38669.3343908784008818&     -2.8617626711234587&     -0.0000430021755164\\
     12.0942471976049308& -38672.1961935693980195& -38669.3343908692986588&     -2.8617626613784495&     -0.0000400387216359\\
     12.2832198100675072& -38672.1961908501034486& -38669.3343908703973284&     -2.8617626521710768&     -0.0000373275324819\\
     12.4721924225300835& -38672.1961883582989685& -38669.3343908699971507&     -2.8617626434351822&     -0.0000348448666045\\
     12.6611650349926617& -38672.1961860805968172& -38669.3343908758033649&     -2.8617626351387764&     -0.0000325696528307\\
     12.8501376474552380& -38672.1961839864015928& -38669.3343908825991093&     -2.8617626272589485&     -0.0000304765417241\\
     13.0391102599178144& -38672.1961820638025529& -38669.3343908882015967&     -2.8617626197497543&     -0.0000285558489850\\
     13.2280828723803907& -38672.1961802919031470& -38669.3343908961978741&     -2.8617626126177083&     -0.0000267830910161\\
     13.4170554848429671& -38672.1961786559986649& -38669.3343909001996508&     -2.8617626057675776&     -0.0000251500314334\\
     13.6060280973055452& -38672.1961771447022329& -38669.3343909024988534&     -2.8617625992344284&     -0.0000236429696088\\
     13.7950007097681215& -38672.1961757442986709& -38669.3343909058967256&     -2.8617625929420685&     -0.0000222454618779\\
     13.9839733222306997& -38672.1961744438012829& -38669.3343909062969033&     -2.8617625868556158&     -0.0000209506470128\\
     14.1729459346932760& -38672.1961732398995082& -38669.3343909057002747&     -2.8617625809371336&     -0.0000197532644961\\
     14.3619185471558524& -38672.1961721158004366& -38669.3343908999013365&     -2.8617625751330085&     -0.0000186407633009\\
     14.5508911596184305& -38672.1961710676987423& -38669.3343908941023983&     -2.8617625694287065&     -0.0000176041648956\\
     14.7398637720810068& -38672.1961700898973504& -38669.3343908880997333&     -2.8617625637976833&     -0.0000166379977600\\
     14.9288363845435850& -38672.1961691788019380& -38669.3343908770984854&     -2.8617625581831256&     -0.0000157435206347\\
     15.1178089970061613& -38672.1961683226982132& -38669.3343908663009643&     -2.8617625526179693&     -0.0000149037805386\\
     15.3067816094687377& -38672.1961675257989555& -38669.3343908555980306&     -2.8617625470572143&     -0.0000141231430462\\
     15.4957542219313140& -38672.1961667751966161& -38669.3343908437964274&     -2.8617625415333974&     -0.0000133898647618\\
     15.6847268343938939& -38672.1961660715969629& -38669.3343908293973072&     -2.8617625360242647&     -0.0000127061721287\\
     15.8736994468564703& -38672.1961654148035450& -38669.3343908168972121&     -2.8617625305584427&     -0.0000120673503261\\
     16.0626720593190484& -38672.1961647927018930& -38669.3343908039023518&     -2.8617625251491550&     -0.0000114636495709\\
     17.0075351216319319& -38672.1961622031012666& -38669.3343907332964591&     -2.8617624993622708&     -0.0000089704408310\\
     17.9523981839448155& -38672.1961602758019581& -38669.3343906756999786&     -2.8617624771033108&     -0.0000071229951573\\
     18.8972612462577025& -38672.1961588236008538& -38669.3343906332011102&     -2.8617624594735531&     -0.0000057309225667\\
     20.7869873708834731& -38672.1961568611004623& -38669.3343905860019731&     -2.8617624372381201&     -0.0000038378639147\\
     24.5664396201350144& -38672.1961549055995420& -38669.3343905661968165&     -2.8617624228469998&     -0.0000019165527192\\
     30.2356179940123226& -38672.1961537999013672& -38669.3343905608999194&     -2.8617624206385144&     -0.0000008183633327\\
     37.7945224925154051& -38672.1961533115027123& -38669.3343905611982336&     -2.8617624205859236&     -0.0000003297172952\\
     47.2431531156442546& -38672.1961531144988840& -38669.3343905610963702&     -2.8617624205856735&     -0.0000001328153303\\
     58.5815098633988782& -38672.1961530371991103& -38669.3343905611982336&     -2.8617624205857228&     -0.0000000554136932\\
     69.9198666111534948& -38672.1961530093976762& -38669.3343905603978783&     -2.8617624205858161&     -0.0000000284126145\\
     75.5890449850308102& -38672.1961530026019318& -38669.3343905607980560&     -2.8617624205856895&     -0.0000000212166924\\
\hline\hline
\end{tabular}
}
\end{table}

\begin{table}[htp]
\centering
\caption{Calculated MRCI interaction potential E$_{Rf^+-He}$, E$_{Rf^+}$ and E$_{He}$ (in atomic units) of the Rf$^+$-He, Rf$^+$-Gh and Gh-He systems, respectively, as function of the interatomic distance d (in atomic unit) obtained for the metastable state $^4$F$_{3/2}$ (7s$^1$6d$^2$) , $\Omega$ = 1/2, together with the corrected potential $V$ (in atomic unit). Gh represents a ghost atom without charge but carrying the the full basis sets of the He or Rf element for the counterpoise energy correction.}
\label{table6}
\resizebox{\textwidth}{!}{
\begin{tabular}{ l  c  c  c c}
\hline\hline
      d &                  E$_{Rf^+-He}$ &          E$_{Rf^+}$ &                 E$_{He}$ &              $V$ corrected\\
\hline
      3.7794522492515403& -38672.1369902114965953& -38669.2942588239966426&     -2.8617634208099165&      0.0190320333131240\\
      3.9684248617141176& -38672.1417277915970772& -38669.2941956927970750&     -2.8617633903946817&      0.0142312915922957\\
      4.1573974741766948& -38672.1450593154004309& -38669.2941467339987867&     -2.8617633625674253&      0.0108507811673917\\
      4.3463700866392712& -38672.1475057275965810& -38669.2941084722988307&     -2.8617633464309575&      0.0083660911332117\\
      4.5353426991018484& -38672.1493745046973345& -38669.2940787806001026&     -2.8617633384086605&      0.0064676143083489\\
      4.7243153115644256& -38672.1508439508033916& -38669.2940557926995098&     -2.8617633277880126&      0.0049751696860767\\
      4.9132879240270029& -38672.1520171907977783& -38669.2940376433034544&     -2.8617633116251135&      0.0037837641284568\\
      5.1022605364895801& -38672.1529566894023446& -38669.2940227145008976&     -2.8617632926781709&      0.0028293177747400\\
      5.2912331489521565& -38672.1537045868026325& -38669.2940098682011012&     -2.8617632722363107&      0.0020685536364908\\
      5.4802057614147337& -38672.1542932000011206& -38669.2939984828990418&     -2.8617632500548273&      0.0014685329515487\\
      5.6691783738773109& -38672.1547497096980806& -38669.2939883223007200&     -2.8617632287870181&      0.0010018413886428\\
      5.8581509863398882& -38672.1550978757004486& -38669.2939793508994626&     -2.8617632120003287&      0.0006446871993830\\
      6.0471235988024654& -38672.1553585079018376& -38669.2939716042965301&     -2.8617632005562852&      0.0003762969499803\\
      6.2360962112650418& -38672.1555495755965239& -38669.2939650987027562&     -2.8617631911040649&      0.0001787142100511\\
      6.4250688237276190& -38672.1556862572033424& -38669.2939597828008118&     -2.8617631796995013&      0.0000367052998627\\
      6.6140414361901954& -38672.1557811020029476& -38669.2939555335979094&     -2.8617631642818164&     -0.0000624041203992\\
      6.8030140486527726& -38672.1558442781970371& -38669.2939521919979597&     -2.8617631460244111&     -0.0001289401770919\\
      6.9919866611153498& -38672.1558838860000833& -38669.2939495867030928&     -2.8617631270118795&     -0.0001711722870823\\
      7.1809592735779262& -38672.1559062782034744& -38669.2939475723032956&     -2.8617631085367603&     -0.0001955973639269\\
      7.3699318860405034& -38672.1559163556012209& -38669.2939460259003681&     -2.8617630907357805&     -0.0002072389615932\\
      7.5589044985030807& -38672.1559178526003961& -38669.2939448550023371&     -2.8617630726482073&     -0.0002099249468301\\
      7.7478771109656570& -38672.1559135511997738& -38669.2939439815963851&     -2.8617630536908774&     -0.0002065159133053\\
      7.9368497234282351& -38672.1559055092002382& -38669.2939433458013809&     -2.8617630339319384&     -0.0001991294702748\\
      8.1258223358908115& -38672.1558951944971341& -38669.2939428855024744&     -2.8617630140373813&     -0.0001892949585454\\
      8.3147949483533896& -38672.1558836600015638& -38669.2939425513977767&     -2.8617629947408458&     -0.0001781138635124\\
      8.5037675608159660& -38672.1558716271974845& -38669.2939422999988892&     -2.8617629766384436&     -0.0001663505609031\\
      8.6927401732785423& -38672.1558595910028089& -38669.2939421002010931&     -2.8617629594517791&     -0.0001545313498355\\
      8.8817127857411204& -38672.1558478781007580& -38669.2939419325994095&     -2.8617629426016649&     -0.0001430029005860\\
      9.0706853982036968& -38672.1558366981989820& -38669.2939417801972013&     -2.8617629253329113&     -0.0001319926668657\\
      9.2596580106662749& -38672.1558261751997634& -38669.2939416445005918&     -2.8617629071461779&     -0.0001216235541506\\
      9.4486306231288513& -38672.1558163734007394& -38669.2939415208966238&     -2.8617628877633070&     -0.0001119647422456\\
      9.6376032355914276& -38672.1558073100022739& -38669.2939414063002914&     -2.8617628672590496&     -0.0001030364437611\\
      9.8265758480540057& -38672.1557989787979750& -38669.2939413022031658&     -2.8617628460341935&     -0.0000948305605561\\
     10.0155484605165821& -38672.1557913482974982& -38669.2939412085979711&     -2.8617628246670970&     -0.0000873150347616\\
     10.2045210729791602& -38672.1557843784030410& -38669.2939411201004987&     -2.8617628037739462&     -0.0000804545270512\\
     10.3934936854417366& -38672.1557780259972787& -38669.2939410417020554&     -2.8617627837647297&     -0.0000742005286156\\
     10.5824662979043129& -38672.1557722345023649& -38669.2939409651007736&     -2.8617627650885922&     -0.0000685043123667\\
     10.7714389103668911& -38672.1557669618996442& -38669.2939408949969220&     -2.8617627478073784&     -0.0000633190938970\\
     10.9604115228294674& -38672.1557621590982308& -38669.2939408279999043&     -2.8617627320573749&     -0.0000585990419495\\
     11.1493841352920455& -38672.1557577866988140& -38669.2939407680969452&     -2.8617627177132672&     -0.0000543008864042\\
     11.3383567477546219& -38672.1557538023989764& -38669.2939407153971842&     -2.8617627046278766&     -0.0000503823757754\\
     11.5273293602171982& -38672.1557501710994984& -38669.2939406695004436&     -2.8617626926173529&     -0.0000468089783681\\
     11.7163019726797764& -38672.1557468601968139& -38669.2939406326986500&     -2.8617626814922730&     -0.0000435460024164\\
     11.9052745851423527& -38672.1557438392992481& -38669.2939406059012981&     -2.8617626711234587&     -0.0000405622777180\\
     12.0942471976049308& -38672.1557410776003962& -38669.2939405845027068&     -2.8617626613784495&     -0.0000378317199647\\
     12.2832198100675072& -38672.1557385560008697& -38669.2939405755969347&     -2.8617626521710768&     -0.0000353282302967\\
     12.4721924225300835& -38672.1557362417006516& -38669.2939405678989715&     -2.8617626434351822&     -0.0000330303664668\\
     12.6611650349926617& -38672.1557341239022207& -38669.2939405686993268&     -2.8617626351387764&     -0.0000309200622723\\
     12.8501376474552380& -38672.1557321739965118& -38669.2939405716970214&     -2.8617626272589485&     -0.0000289750387310\\
     13.0391102599178144& -38672.1557303816007334& -38669.2939405751967570&     -2.8617626197497543&     -0.0000271866520052\\
     13.2280828723803907& -38672.1557287274990813& -38669.2939405819997774&     -2.8617626126177083&     -0.0000255328850471\\
     13.4170554848429671& -38672.1557271981000667& -38669.2939405853976496&     -2.8617626057675776&     -0.0000240069348365\\
     13.6060280973055452& -38672.1557257828972070& -38669.2939405873985379&     -2.8617625992344284&     -0.0000225962648983\\
     13.7950007097681215& -38672.1557244697032729& -38669.2939405907964101&     -2.8617625929420685&     -0.0000212859667954\\
     13.9839733222306997& -38672.1557232479972299& -38669.2939405911965878&     -2.8617625868556158&     -0.0000200699432753\\
     14.1729459346932760& -38672.1557221156981541& -38669.2939405908982735&     -2.8617625809371336&     -0.0000189438651432\\
     14.3619185471558524& -38672.1557210565006244& -38669.2939405853030621&     -2.8617625751330085&     -0.0000178960617632\\
     14.5508911596184305& -38672.1557200675015338& -38669.2939405798970256&     -2.8617625694287065&     -0.0000169181730598\\
     14.7398637720810068& -38672.1557191433021217& -38669.2939405739016365&     -2.8617625637976833&     -0.0000160056006280\\
     14.9288363845435850& -38672.1557182812975952& -38669.2939405634970171&     -2.8617625581831256&     -0.0000151596177602\\
     15.1178089970061613& -38672.1557174698973540& -38669.2939405530996737&     -2.8617625526179693&     -0.0000143641809700\\
     15.3067816094687377& -38672.1557167139981175& -38669.2939405430006445&     -2.8617625470572143&     -0.0000136239395943\\
     15.4957542219313140& -38672.1557160009979270& -38669.2939405318029458&     -2.8617625415333974&     -0.0000129276595544\\
     15.6847268343938939& -38672.1557153320027282& -38669.2939405178985908&     -2.8617625360242647&     -0.0000122780766105\\
     15.8736994468564703& -38672.1557147068015183& -38669.2939405060969875&     -2.8617625305584427&     -0.0000116701485240\\
     16.0626720593190484& -38672.1557141137964209& -38669.2939404938006192&     -2.8617625251491550&     -0.0000110948458314\\
     17.0075351216319319& -38672.1557116387994029& -38669.2939404268036014&     -2.8617624993622708&     -0.0000087126318249\\
     17.9523981839448155& -38672.1557097893019090& -38669.2939403726995806&     -2.8617624771033108&     -0.0000069394955062\\
     18.8972612462577025& -38672.1557083906009211& -38669.2939403334021335&     -2.8617624594735531&     -0.0000055977216107\\
     20.7869873708834731& -38672.1557064922017162& -38669.2939402904012240&     -2.8617624372381201&     -0.0000037645659177\\
     24.5664396201350144& -38672.1557045859008213& -38669.2939402721021906&     -2.8617624228469998&     -0.0000018909486244\\
     30.2356179940123226& -38672.1557034990983084& -38669.2939402667034301&     -2.8617624206385144&     -0.0000008117567631\\
     37.7945224925154051& -38672.1557030159019632& -38669.2939402670017444&     -2.8617624205859236&     -0.0000003283130354\\
     47.2431531156442546& -38672.1557028200986679& -38669.2939402668998810&     -2.8617624205856735&     -0.0000001326116035\\
     58.5815098633988782& -38672.1557027430972084& -38669.2939402671036078&     -2.8617624205857228&     -0.0000000554064172\\
     69.9198666111534948& -38672.1557027153030504& -38669.2939402663032524&     -2.8617624205858161&     -0.0000000284126145\\
     75.5890449850308102& -38672.1557027085000300& -38669.2939402667034301&     -2.8617624205856895&     -0.0000000212094164\\
\hline\hline
\end{tabular}
}
\end{table}

\begin{table}[htp]
\centering
\caption{Calculated MRCI interaction potential E$_{Rf^+-He}$, E$_{Rf^+}$ and E$_{He}$ (in atomic units) of the Rf$^+$-He, Rf$^+$-Gh and Gh-He systems, respectively, as function of the interatomic distance d (in atomic unit) obtained for the metastable state $^4$F$_{3/2}$ (7s$^1$6d$^2$) , $\Omega$ = 3/2, together with the corrected potential $V$ (in atomic unit). Gh represents a ghost atom without charge but carrying the the full basis sets of the He or Rf element for the counterpoise energy correction.}
\label{table7}
\resizebox{\textwidth}{!}{
\begin{tabular}{ l  c  c  c c}
\hline\hline
      d &                  E$_{Rf^+-He}$ &          E$_{Rf^+}$ &                 E$_{He}$ &              $V$ corrected\\
\hline
      3.7794522492515403& -38672.1393667918018764& -38669.2941926498970133&     -2.8617634208099165&      0.0165892789082136\\
      3.9684248617141176& -38672.1438247999030864& -38669.2941394666995620&     -2.8617633903946817&      0.0120780571887735\\
      4.1573974741766948& -38672.1468629149021581& -38669.2940990486968076&     -2.8617633625674253&      0.0089994963636855\\
      4.3463700866392712& -38672.1490230181007064& -38669.2940680417013937&     -2.8617633464309575&      0.0068083700316492\\
      4.5353426991018484& -38672.1506292324993410& -38669.2940443892002804&     -2.8617633384086605&      0.0051784951065201\\
      4.7243153115644256& -38672.1518685107002966& -38669.2940263761993265&     -2.8617633277880126&      0.0039211932889884\\
      4.9132879240270029& -38672.1528460336994613& -38669.2940123759035487&     -2.8617633116251135&      0.0029296538268682\\
      5.1022605364895801& -38672.1536223633011105& -38669.2940010192032787&     -2.8617632926781709&      0.0021419485783554\\
      5.2912331489521565& -38672.1542359896993730& -38669.2939913517984678&     -2.8617632722363107&      0.0015186343371170\\
      5.4802057614147337& -38672.1547151561971987& -38669.2939828482994926&     -2.8617632500548273&      0.0010309421559214\\
      5.6691783738773109& -38672.1550831567001296& -38669.2939752992024296&     -2.8617632287870181&      0.0006553712883033\\
      5.8581509863398882& -38672.1553602649000823& -38669.2939686604004237&     -2.8617632120003287&      0.0003716075007105\\
      6.0471235988024654& -38672.1555642465027631& -38669.2939629531974788&     -2.8617632005562852&      0.0001619072500034\\
      6.2360962112650418& -38672.1557104497987893& -38669.2939581902974169&     -2.8617631911040649&      0.0000109316024464\\
      6.4250688237276190& -38672.1558118225002545& -38669.2939543335014605&     -2.8617631796995013&     -0.0000943092964008\\
      6.6140414361901954& -38672.1558790358976694& -38669.2939512859011302&     -2.8617631642818164&     -0.0001645857119001\\
      6.8030140486527726& -38672.1559206934980466& -38669.2939489205964492&     -2.8617631460244111&     -0.0002086268796120\\
      6.9919866611153498& -38672.1559436066017952& -38669.2939470975034055&     -2.8617631270118795&     -0.0002333820884814\\
      7.1809592735779262& -38672.1559530774029554& -38669.2939456983003765&     -2.8617631085367603&     -0.0002442705663270\\
      7.3699318860405034& -38672.1559531622988288& -38669.2939446235977812&     -2.8617630907357805&     -0.0002454479617882\\
      7.5589044985030807& -38672.1559469264029758& -38669.2939438020985108&     -2.8617630726482073&     -0.0002400516532362\\
      7.7478771109656570& -38672.1559366298024543& -38669.2939431775012054&     -2.8617630536908774&     -0.0002303986111656\\
      7.9368497234282351& -38672.1559239284979412& -38669.2939427121964400&     -2.8617630339319384&     -0.0002181823729188\\
      8.1258223358908115& -38672.1559099840014824& -38669.2939423659991007&     -2.8617630140373813&     -0.0002046039662673\\
      8.3147949483533896& -38672.1558956161970855& -38669.2939421092014527&     -2.8617629947408458&     -0.0001905122553580\\
      8.5037675608159660& -38672.1558813680021558& -38669.2939419133981573&     -2.8617629766384436&     -0.0001764779663063\\
      8.6927401732785423& -38672.1558675960986875& -38669.2939417582019814&     -2.8617629594517791&     -0.0001628784448258\\
      8.8817127857411204& -38672.1558545196967316& -38669.2939416298031574&     -2.8617629426016649&     -0.0001499472928117\\
      9.0706853982036968& -38672.1558422638991033& -38669.2939415135988384&     -2.8617629253329113&     -0.0001378249653499\\
      9.2596580106662749& -38672.1558308863968705& -38669.2939414111024234&     -2.8617629071461779&     -0.0001265681494260\\
      9.4486306231288513& -38672.1558203999011312& -38669.2939413174026413&     -2.8617628877633070&     -0.0001161947366199\\
      9.6376032355914276& -38672.1558107824967010& -38669.2939412284031278&     -2.8617628672590496&     -0.0001066868353519\\
      9.8265758480540057& -38672.1558019977965159& -38669.2939411457991810&     -2.8617628460341935&     -0.0000980059630820\\
     10.0155484605165821& -38672.1557939920967328& -38669.2939410696999403&     -2.8617628246670970&     -0.0000900977320271\\
     10.2045210729791602& -38672.1557867087976774& -38669.2939409957034513&     -2.8617628037739462&     -0.0000829093187349\\
     10.3934936854417366& -38672.1557800922018941& -38669.2939409294995130&     -2.8617627837647297&     -0.0000763789357734\\
     10.5824662979043129& -38672.1557740763964830& -38669.2939408637976157&     -2.8617627650885922&     -0.0000704475096427\\
     10.7714389103668911& -38672.1557686119995196& -38669.2939408038000693&     -2.8617627478073784&     -0.0000650603906251\\
     10.9604115228294674& -38672.1557636442012154& -38669.2939407464000396&     -2.8617627320573749&     -0.0000601657447987\\
     11.1493841352920455& -38672.1557591286982642& -38669.2939406955993036&     -2.8617627177132672&     -0.0000557153834961\\
     11.3383567477546219& -38672.1557550193974748& -38669.2939406513032736&     -2.8617627046278766&     -0.0000516634681844\\
     11.5273293602171982& -38672.1557512779982062& -38669.2939406130026327&     -2.8617626926173529&     -0.0000479723748867\\
     11.7163019726797764& -38672.1557478691975120& -38669.2939405826982693&     -2.8617626814922730&     -0.0000446050034952\\
     11.9052745851423527& -38672.1557447606028290& -38669.2939405610013637&     -2.8617626711234587&     -0.0000415284812334\\
     12.0942471976049308& -38672.1557419197997660& -38669.2939405435026856&     -2.8617626613784495&     -0.0000387149193557\\
     12.2832198100675072& -38672.1557393264010898& -38669.2939405373035697&     -2.8617626521710768&     -0.0000361369238817\\
     12.4721924225300835& -38672.1557369466972887& -38669.2939405311990413&     -2.8617626434351822&     -0.0000337720630341\\
     12.6611650349926617& -38672.1557347690977622& -38669.2939405326978886&     -2.8617626351387764&     -0.0000316012592521\\
     12.8501376474552380& -38672.1557327647024067& -38669.2939405357974465&     -2.8617626272589485&     -0.0000296016442007\\
     13.0391102599178144& -38672.1557309224008350& -38669.2939405388970044&     -2.8617626197497543&     -0.0000277637518593\\
     13.2280828723803907& -38672.1557292228026199& -38669.2939405451979837&     -2.8617626126177083&     -0.0000260649903794\\
     13.4170554848429671& -38672.1557276519015431& -38669.2939405479992274&     -2.8617626057675776&     -0.0000244981347350\\
     13.6060280973055452& -38672.1557261990019470& -38669.2939405495999381&     -2.8617625992344284&     -0.0000230501682381\\
     13.7950007097681215& -38672.1557248515964602& -38669.2939405528013594&     -2.8617625929420685&     -0.0000217058550334\\
     13.9839733222306997& -38672.1557235989021137& -38669.2939405532015371&     -2.8617625868556158&     -0.0000204588432098\\
     14.1729459346932760& -38672.1557224382995628& -38669.2939405532015371&     -2.8617625809371336&     -0.0000193041632883\\
     14.3619185471558524& -38672.1557213537016651& -38669.2939405481010908&     -2.8617625751330085&     -0.0000182304647751\\
     14.5508911596184305& -38672.1557203416014090& -38669.2939405434008222&     -2.8617625694287065&     -0.0000172287691385\\
     14.7398637720810068& -38672.1557193964981707& -38669.2939405384968268&     -2.8617625637976833&     -0.0000162942014867\\
     14.9288363845435850& -38672.1557185155033949& -38669.2939405293000164&     -2.8617625581831256&     -0.0000154280205606\\
     15.1178089970061613& -38672.1557176869973773& -38669.2939405201977934&     -2.8617625526179693&     -0.0000146141828736\\
     15.3067816094687377& -38672.1557169154984877& -38669.2939405115030240&     -2.8617625470572143&     -0.0000138569375849\\
     15.4957542219313140& -38672.1557161883029039& -38669.2939405018987600&     -2.8617625415333974&     -0.0000131448687171\\
     15.6847268343938939& -38672.1557155065020197& -38669.2939404896023916&     -2.8617625360242647&     -0.0000124808721012\\
     15.8736994468564703& -38672.1557148696010699& -38669.2939404795033624&     -2.8617625305584427&     -0.0000118595417007\\
     16.0626720593190484& -38672.1557142660021782& -38669.2939404689022922&     -2.8617625251491550&     -0.0000112719499157\\
     17.0075351216319319& -38672.1557117496995488& -38669.2939404099015519&     -2.8617624993622708&     -0.0000088404340204\\
     17.9523981839448155& -38672.1557098726989352& -38669.2939403620985104&     -2.8617624771033108&     -0.0000070334936026\\
     18.8972612462577025& -38672.1557084544983809& -38669.2939403272030177&     -2.8617624594735531&     -0.0000056678181863\\
     20.7869873708834731& -38672.1557065309025347& -38669.2939402882984723&     -2.8617624372381201&     -0.0000038053694880\\
     24.5664396201350144& -38672.1557046016969252& -38669.2939402717966004&     -2.8617624228469998&     -0.0000019070503186\\
     30.2356179940123226& -38672.1557035043006181& -38669.2939402667034301&     -2.8617624206385144&     -0.0000008169590728\\
     37.7945224925154051& -38672.1557030175972614& -38669.2939402670017444&     -2.8617624205859236&     -0.0000003300083335\\
     47.2431531156442546& -38672.1557028207025724& -38669.2939402668998810&     -2.8617624205856735&     -0.0000001332155080\\
     58.5815098633988782& -38672.1557027433009353& -38669.2939402671036078&     -2.8617624205857228&     -0.0000000556101440\\
     69.9198666111534948& -38672.1557027153976378& -38669.2939402663032524&     -2.8617624205858161&     -0.0000000285072019\\
     75.5890449850308102& -38672.1557027086018934& -38669.2939402667034301&     -2.8617624205856895&     -0.0000000213112799\\
\hline\hline
\end{tabular}
}
\end{table}

\begin{table}[t]
\caption{Calculated equilibrium distances $d$$_{min}$ (in {\AA}) and dissociation energies (in cm$^{-1}$) for the Rf$^+$-He derived from the MRCI interaction potentials with ($V$) and without ($V$ correct.) counterpoise energy correction for the electronic ground state, $^2$D$_{3/2}$ (6d$^1$7s$^2$), the low-lying excited state, $^2$D$_{5/2}$ (6d$^1$7s$^2$) and the metastable state, $^4$F$_{3/2}$ (6d$^2$7s$^1$), with their $\Omega$ projections.}
\centering
\label{table8}
\begin{tabular}{ll|cc|cc}
\hline \hline
&                                        & \multicolumn{2}{c|}{$V$ correct.}  & \multicolumn{2}{c}{$V$} \\
&$^{(2S+1)}$L$_J$              $\Omega$  & $d$$_{min}$ & $D$$_e$              &$d$$_{min}$ & $D$$_e$ \\ 
\hline 
\multirow{2}{*}{$^2$D$_{3/2}$} & 1/2      & 4.101       & 43.817               & 4.089      &44.895 \\
                               & 3/2      & 4.065       & 47.875               & 4.059      &48.456 \\ 
\multirow{3}{*}{$^2$D$_{5/2}$} & 1/2      & 4.193       & 39.561               & 4.178      &40.721 \\
                               & 3/2      & 3.942       & 52.794               & 3.929      &53.889 \\
                               & 5/2      & 4.065       & 48.503               & 4.062      &49.029 \\ 
\multirow{2}{*}{$^4$F$_{3/2}$} & 1/2      & 3.987       & 46.076               & 3.971      &47.275 \\ 
                               & 3/2      & 3.864       & 53.967               & 3.849      &55.167 \\  
\hline 
\hline
\end{tabular}
\end{table}

\begin{figure}[h]
\centering
\includegraphics[width=0.72\textwidth]{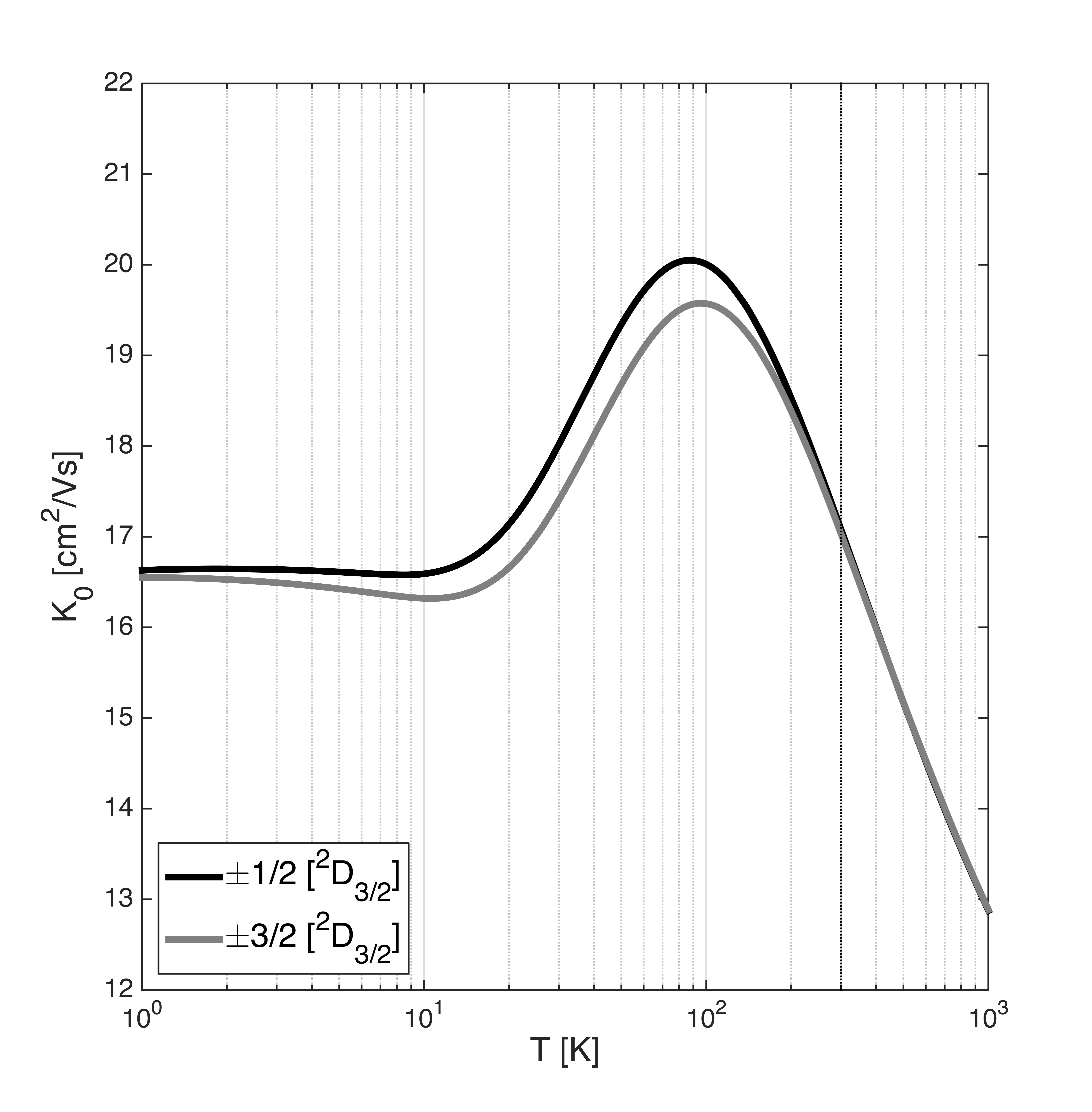}
\caption{Reduced zero-field mobilities of the Rf$^+$-He system in the ground $^2$D$_{3/2}$ (7s$^2$6d$^1$) state: comparison between the results obtained from the MRCI interaction potentials for the two components $\pm$1/2 and $\pm$3/2. Note that in the maintext, the ion-mobility trace for $^2$D$_{3/2}$ (7s$^2$6d$^1$)  is an average between the represented data.
}
\label{figure1}
\end{figure} 

\begin{figure}[h]
\centering
\includegraphics[width=0.72\textwidth]{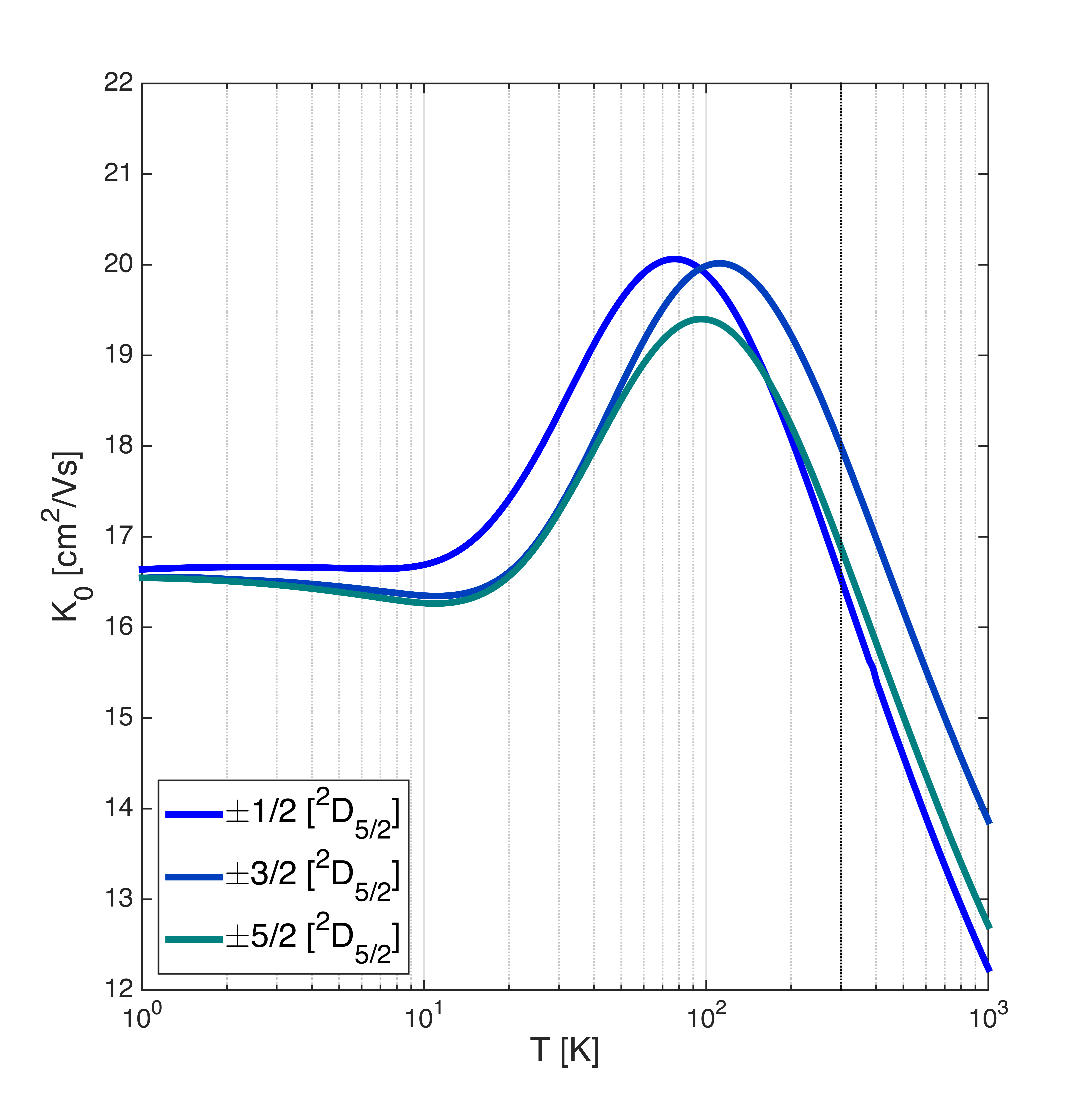}
\caption{Reduced zero-field mobilities of the Rf$^+$-He system in the low-lying $^2$D$_{5/2}$ (7s$^2$6d$^1$) state: comparison between the results obtained from the MRCI interaction potentials for the three components $\pm$1/2, $\pm$3/2 and $\pm$5/2. Note that in the maintext, the ion-mobility trace for $^2$D$_{5/2}$ (7s$^2$6d$^1$)  is an average between the represented data. 
}
\label{figure2}
\end{figure}

\begin{figure}[h]
\centering
\includegraphics[width=0.72\textwidth]{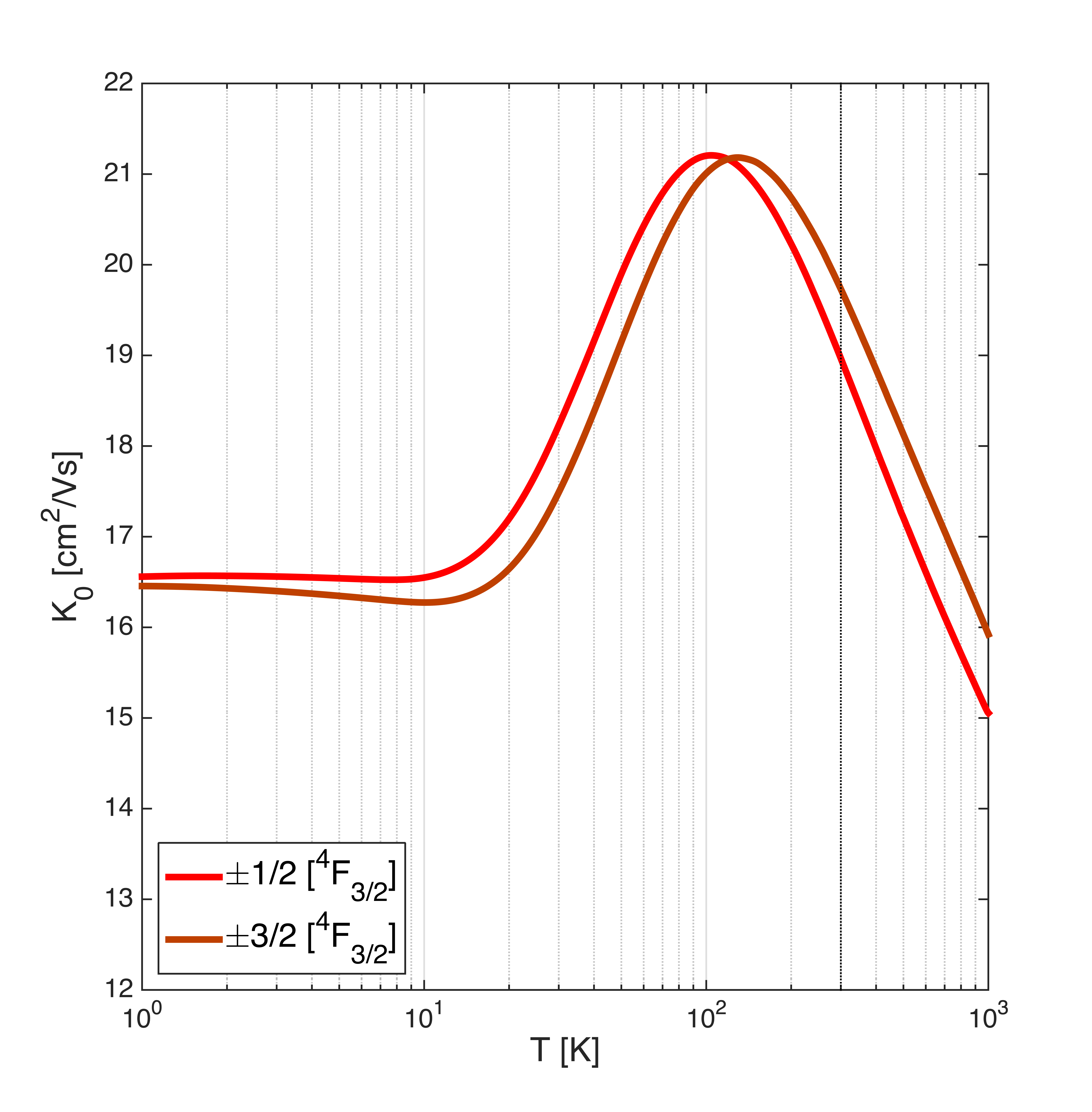}
\caption{Reduced zero-field mobilities of the Rf$^+$-He system in the metastable $^4$F$_{3/2}$ (7s$^1$6d$^2$) state: comparison between the results obtained from the MRCI interaction potentials for the two components $\pm$1/2 and $\pm$3/2. Note that in the maintext, the ion-mobility trace for $^4$F$_{3/2}$ (7s$^1$6d$^1$) is an average between the represented data. 
}
\label{figure3}
\end{figure} 

\begin{figure}[h]
\centering
\includegraphics[width=0.72\textwidth]{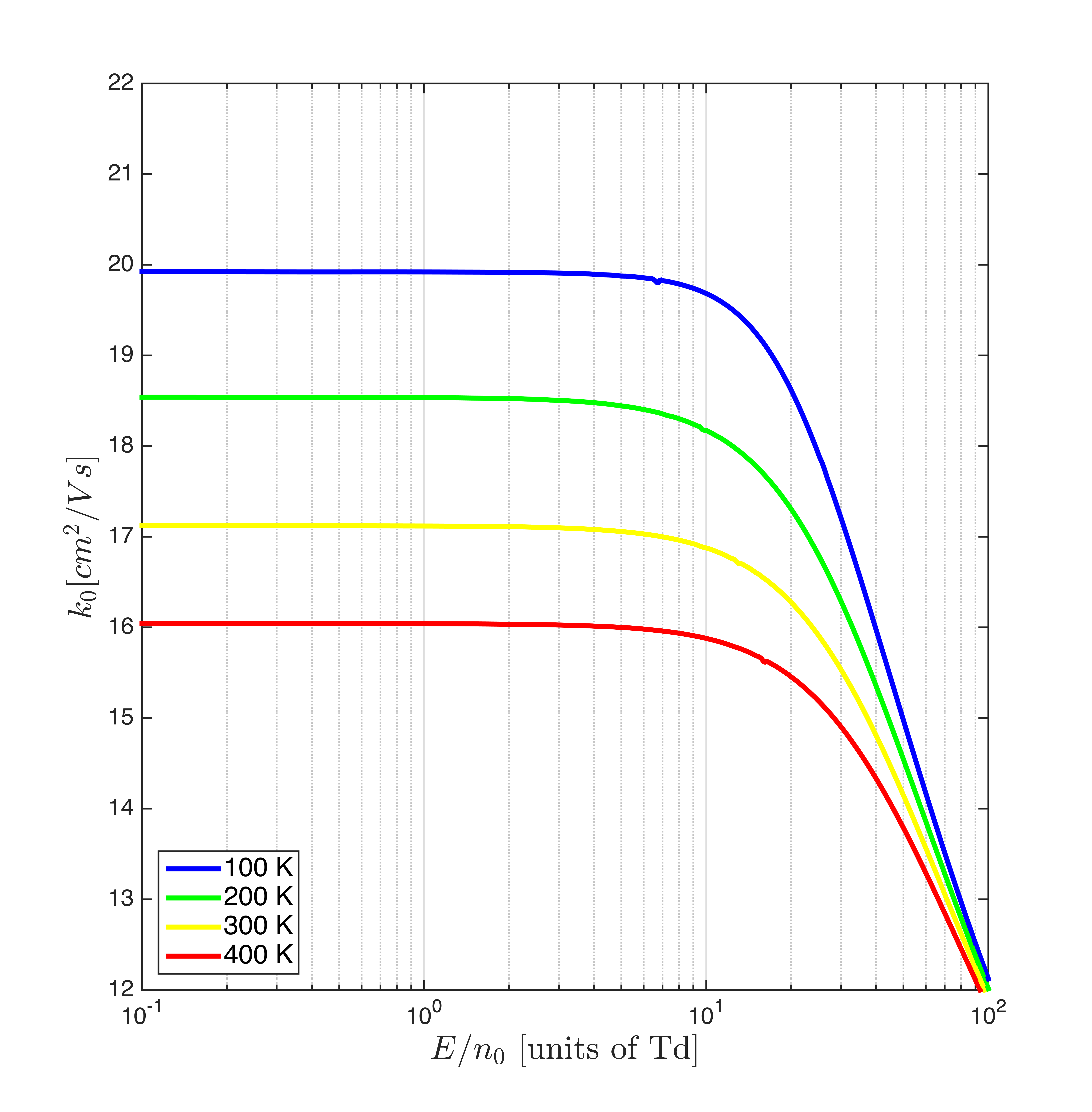}
\caption{Reduced mobilities as function of $E/n_0$ at selected temperatures of the Rf$^+$-He system in the ground $^2$D$_{3/2}$ (7s$^2$6d$^1$) state: comparison between the results obtained from the MRCI interaction potentials for the two components $\pm$1/2 (solid line) and $\pm$3/2 (dashed line). Note that in the maintext, the ion-mobility trace for $^2$D$_{3/2}$ (7s$^2$6d$^1$)  is an average between the represented data at selective temperature.
}
\label{figure4}
\end{figure}

\begin{figure}[h]
\centering
\includegraphics[width=0.72\textwidth]{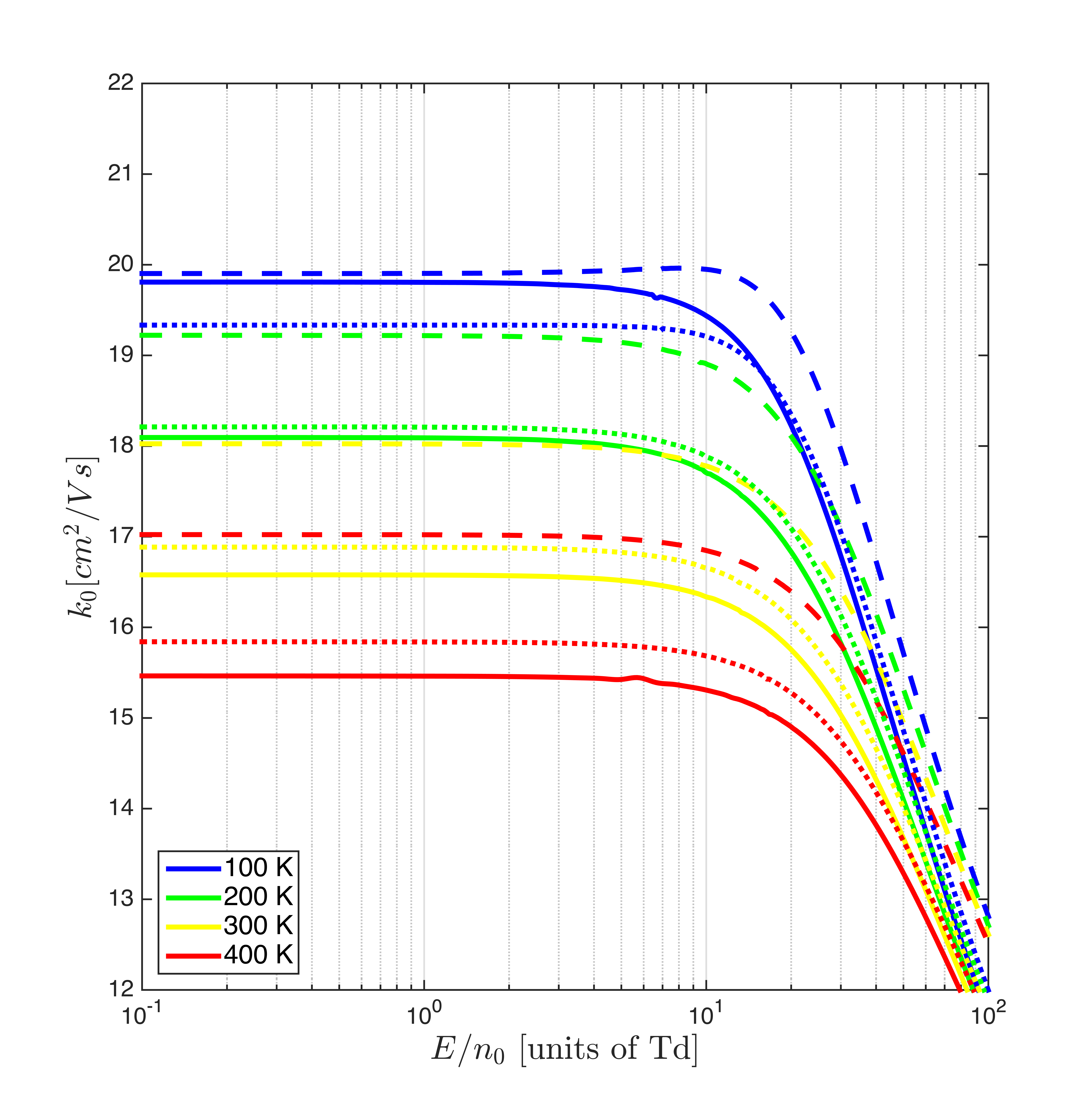}
\caption{Reduced mobilities as function of $E/n_0$ at selected temperatures of the Rf$^+$-He system in the low-lying $^2$D$_{5/2}$ (7s$^2$6d$^1$) state: comparison between the results obtained from the MRCI interaction potentials for the three components $\pm$1/2 (solid line), $\pm$3/2 (dashed line) and $\pm$5/2 (dotted line). Note that in the maintext, the ion-mobility trace for $^2$D$_{5/2}$ (7s$^2$6d$^1$) is an average between the represented data at selective temperature.
}
\label{figure5}
\end{figure}

\begin{figure}[h]
\centering
\includegraphics[width=0.72\textwidth]{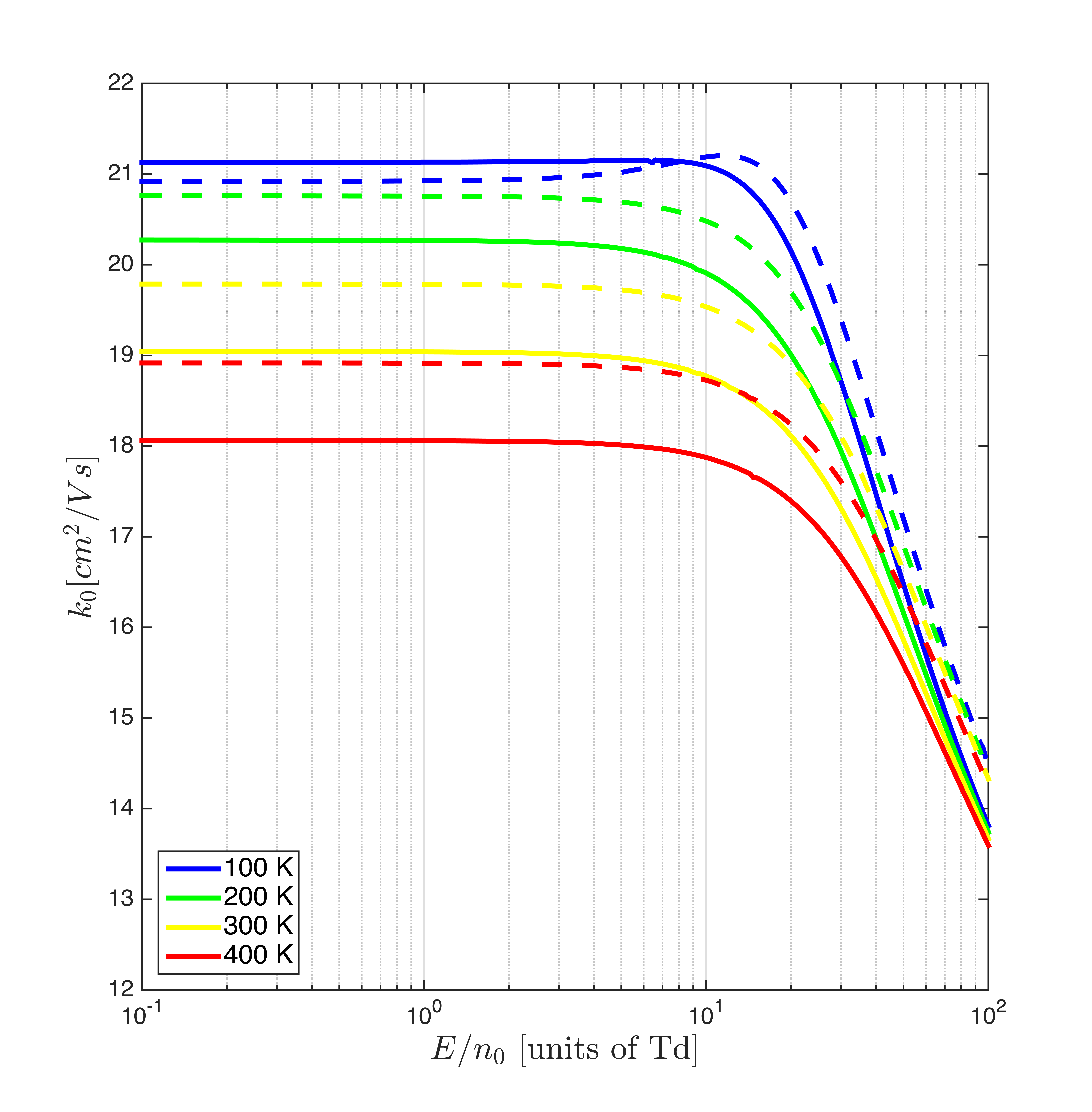}
\caption{Reduced mobilities as function of $E/n_0$ at selected temperatures of the Rf$^+$-He system in the metastable $^4$F$_{3/2}$ (7s$^1$6d$^2$) state: comparison between the results obtained from the MRCI interaction potentials for the two components $\pm$1/2 (solid line) and $\pm$3/2 (dashed line). Note that in the maintext, the ion-mobility trace for $^4$F$_{3/2}$ (7s$^1$6d$^2$)  is an average between the represented data at selective temperature.
}
\label{figure6}
\end{figure}